\documentclass[aps,prb,nopacs,superscriptaddress,longbibliography,reprint]{revtex4-1}
\usepackage[breaklinks=true,colorlinks,citecolor=blue,linkcolor=blue,urlcolor=blue]{hyperref}
\usepackage{graphicx,epsfig}% Include figure files
\usepackage{amsmath}% bold math
\usepackage{subfigure}

\usepackage[table]{xcolor}
\renewcommand{\BibitemShut}[1]{}
\usepackage{siunitx}
\def\be{\begin{equation}}
	\def\ee{\end{equation}}

\def\bea{\begin{eqnarray}}
	\def\eea{\end{eqnarray}}

\begin{document}
	
	%preprint{APS/123-QED}

	%\title{Ultra-low-power steep-slope atomic channel transistors based on laterally confined monolayer MoSi$_2$N$_4$}
	
	%\title{Intrinsic cold contact effect in laterally confined MA2Z2 monolayers for steep-slope transistors}% Force line breaks with \\
	
	%\title{Overcoming Boltzmann’s Tyranny in a Transistor via the Cold Contact Effect in Laterally Confined MA2Z2 Monolayers}
	
	%\title{Overcoming Boltzmann's Tyranny in Transistors with the Cold Contact Effect in Laterally Confined Semiconducting MA$_2$Z$_4$ Monolayers}
	
	%\title{Laterally confined semiconducting MA$_2$Z$_4$ monolayers for next-generation low-power logic electronics}
	\title{Designing power efficient transistors using narrow bandwidth materials from the MA$_2$Z$_4$ monolayer series}

	\author{Keshari Nandan}
	\email{keshari@iitk.ac.in}
	\affiliation{Department of Electrical Engineering, Indian Institute of Technology Kanpur, India, 208016.}
	
	\author{Somnath Bhowmick}
	\email{bsomnath@iitk.ac.in}
	\affiliation{Department of Materials Science and Engineering, Indian Institute of Technology Kanpur, India, 208016.}
	
	\author{Yogesh S. Chauhan}
	\email{chauhan@iitk.ac.in}
	\affiliation{Department of Electrical Engineering, Indian Institute of Technology Kanpur, India, 208016.}
	
	\author{Amit Agarwal}
	\email{amitag@iitk.ac.in}
	\affiliation{Department of Physics, Indian Institute of Technology Kanpur, India, 208016.}
	
	\date{\today}% It is always \today, today,
	
	\begin{abstract}
		
		The subthreshold leakage current in transistors has become a critical limiting factor for realizing ultra-low-power transistors. The leakage current is predominantly dictated by the long thermal tail of the charge carriers. We propose a solution to this problem by using narrow bandwidth semiconductors for limiting the thermionic leakage current by filtering out the high energy carriers.
		We specifically demonstrate this solution in transistors with laterally confined monolayer MoSi$_2$N$_4$ with different passivation serving as channel material. 
		Remarkably, we find that the proposed narrow bandwidth devices can achieve a large ON/OFF current ratio with an ultra-low-power supply voltage of $\sim$ 0.1 V, even for devices with $\sim$ 5 nm gate length. We also show that several other materials share the unique electronic properties of narrow bandwidth conduction and valance bands in the same series. 
		%This opens up new avenues for effectively tackling the OFF state current leakage and power dissipation problem and realizing ultra-low-power transistors.
		%The subthreshold leakage current in transistors has become a critical limiting factor for realizing ultra low-power transistors. The leakage current is predominantly dictated by the long thermal tail of the charge carriers. We propose a solution to this problem by using narrow bandwidth semiconductors, to limit the thermionic leakage current by filtering out the high energy carriers.
		%We demonstrate this using laterally confined monolayer MoSi$_2$N$_4$, with different passivation, as a channel material in two-dimensional transistors.  Using this method, the subthreshold slope can be reduced to remarkably low values of up to ~20 mV/decade at room temperature. We show that the unique electronic properties of narrow bandwidth conduction and valance bands are also shared by other materials in the same series. This opens up new avenues for realizing ultra low-power transistors and computing.   
	\end{abstract}
	
	\maketitle
	\section{Introduction}
\begin{figure*}[!t]
		\includegraphics[width = 0.999\linewidth]{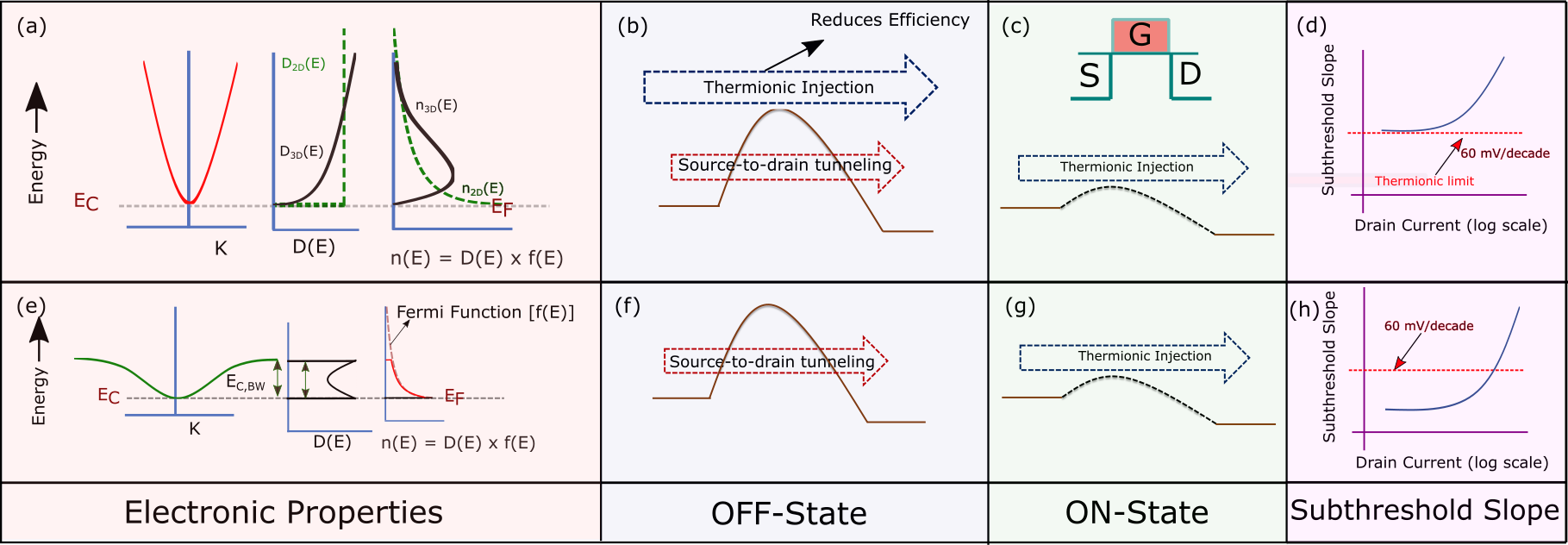}
		\caption{Schematic depicting the idea of ultra-low-power transistors based on narrow bandwidth materials, which suppress the thermionic injection in the OFF state. (a) The energy dispersion, the density of states [D(E)], and electron density [n(E)] with a long thermal tail in conventional 3D and 2D materials. (b) The OFF state has an unwanted leakage current arising from the long thermal tail of the carriers (thermionic injection) and tunneling. (c) The ON state with a lowered potential barrier and the ON current arising from the thermionic injection. (d) The subthreshold slope is limited to 60 mV/decade in conventional transistors of 3-D and 2-D semiconductors due to the long thermal tail. (e) Novel semiconductors with narrow bandwidth, which suppresses the thermal tail of the carriers. (f) The lack of a long thermal tail of carriers suppresses the thermionic injection in the OFF state. This enables subthermionic switching leading to ultra-efficient transistors. (g) The ON state. 
			%with a lowered potential barrier and the ON current arising from the thermionic injection. 
			(h) The subthreshold slope overcomes the thermionic limit due to the short thermal tail in these narrow bandwidth semiconductors.
			\label{FIG_1}}
	\end{figure*}

	\vspace{-4.0mm}
	The main bottleneck in the design of ultra low-power transistors is the rate change of the ON current with the applied gate voltage \cite{Adrian}. This is technically described as the lowering of the subthreshold slope ($SS$), which is a crucial figure of merit determining a transistor's performance. The $SS$ is the smallest change in gate voltage, $V_\mathrm{GS}$ required to change the drain current, $I_\mathrm{DS}$ by one order of magnitude. The dissipation of energy by thermally excited charge carriers is facilitated by a existence of electronic states at high energies, and it sets a 
  natural limit on the subthreshold slope to be $SS < 60$ mV/decade. This is known as the thermionic limit and this thermodynamic dissipation problem is  referred to as Boltzmann's tyranny (see Fig.~\ref{FIG_1}). 
  %
  %Physically, it arises due to the long thermal tail of the charge carriers (see Figure~\ref{FIG_1}). 
 %
	%Overcoming the Boltzmann tyranny is currently one of the biggest challenges in conceptualizing ultra low-power transistors. To solve this problem and overcome Boltzmann's tyranny, different strategies are being used. One strategy is to shorten the thermal tail, with tunnel field-effect transistor (FET)  \cite{Qin_TFET, 4457828, Knoch_PRL} being the most common device. Another strategy is to increase the barrier height's steepness with gate voltages, with Landau's FET being the most common device \cite{Sayeef,adrian_ncfet,Rusu}. However, the tunneling transport limits the ON-state current to very low values in tunnel FETs, and there are several challenges in Landau FETs at both material and device levels. In fact, low hysteresis with the required ON-state current has not been achieved experimentally \cite{Si2018,7409759,7409760,7331265} {in Landau FETs}.  As another solution, the Dirac cone in the electronic band structures of layered materials such as graphene has also been exploited to shorten the thermal tail for subthermionic switching  \cite{qiu2018dirac, Gr_MoS2_DSFET}. Layered materials have also shown great potential for two dimensional transistors \cite{Radisavljevic2011,Roy_Nano, Desai99, Li2020,DT-FET,Liu2020,Pala_1,Nadeem,Li2020,Shen2021,Qiu_CNT_Science, Sun_nano, Wu2022, Zou2020, Lemme2022, carbon, Ganesh_Nano, Butler_Nano, Lin_Nano, Cai_CR, DD_Nano}.
	%
	Overcoming the Boltzmann tyranny by making $SS < 60$ mV/decade is currently one of the biggest challenges in conceptualizing ultra-low-power transistors. 
 
 Different strategies are being used to solve this problem. One strategy is to shorten the thermal tail, with tunnel field-effect transistors (FETs) \cite{Qin_TFET, 4457828, Knoch_PRL} being the most common device. Another strategy is to increase the barrier height's steepness with gate voltages, with Landau's FETs \cite{Sayeef,adrian_ncfet,Rusu} being the most common device. However, the tunneling transport limits the ON-state current to very low values in tunnel FETs, and there are several challenges in Landau FETs at both material and device levels. In fact, low hysteresis with the required ON-state current has not been achieved experimentally \cite{Si2018,7409759,7409760,7331265} in Landau FETs. In recent years, the rise of layered materials \cite{Radisavljevic2011, Roy_Nano, Desai99, Li2020, Wu2022, Zou2020, Lemme2022, carbon, Ganesh_Nano, Butler_Nano, Lin_Nano, Cai_CR, DD_Nano, D2NR02382B, PhysRevB.102.035420, PhysRevB.100.235101, Ghosh_jpcc, PhysRevB.94.205426} has offered more opportunities to conceptualize and design energy-efficient logic devices \cite{DT-FET,Liu2020,Pala_1,Nadeem}. For subthermionic switching, the Dirac cone in the electronic band structures of layered materials like graphene has also been used to shorten the thermal tail \cite{qiu2018dirac, Gr_MoS2_DSFET}.

	Here, we present a solution to the problem of Boltzmann's tyranny, by using narrow bandwidth semiconductors as channel material. The narrow bandwidth of the conduction/valance band naturally restricts the long thermal tail of the charge carriers, reduces the thermionic current in the OFF state [see Figure~\ref{FIG_1} (f)], and solves the problem of Boltzmann's tyranny. We demonstrate that using this solution, the devices can achieve a large ON/OFF current ratio with an ultra-low-power supply voltage of $\sim$ 0.1 V, down to $\sim$ 5 nm gate length. While the concept is very general, we demonstrate this in the laterally confined monolayer of the recently synthesized MoSi$_2$N$_4$ monolayers \cite{Hong}. Monolayer MoSi$_2$N$_4$ is an excellent semiconductor with outstanding physical, mechanical, thermal, electronic, and metal contact properties, which are at par with most other 2-D semiconductors \cite{Hong, Cao, Bafekry_2021,MORTAZAVI2021105716,Jian,Wang2021,9646230}. We assess the transistor performance of differently passivated monolayers of MoSi$_2$N$_4$, by using first principles-based nanoscale device simulations. All single and dual-gate devices show ultra-low-power steep slope transistor characteristics down to short channel lengths. Additionally, we show that other members of the MoSi$_2$N$_4$ family also show these unique electronic properties in their laterally confined structures. This opens up new avenues for exploring 2-D transistors for ultra-power-efficient computing. 
 
	This paper is organized as follows: In Sec. \ref{RD_A}, we describe the origin of low power dissipation and subthermionic switching in transistors with narrow bandwidth source materials. As an example of this, we discuss the electronic properties of passivated MoSi$_2$N$_4$ nano-ribbons, which host narrow bandwidth conduction band in Sec.~\ref{RD_B}. In Sec. \ref{RD_C}, we describe the electrical properties of transistors made from these structures. In Secs. \ref{RD_D} and \ref{RD_E}, we show that several differently terminated MoSi$_2$N$_4$, and other materials in the MoSi$_2$N$_4$ series also host narrow conduction or valance bands. These can also serve as power efficient source or channel materials. Finally, we summarize our findings in Sec. \ref{Conclusion}.
	%%%%%%%%%%%%%%%%%%%%-----Results and Discussion
	\section{ Results and Discussions }\label{RD}
	\vspace{-4.0mm}
	\subsection{Narrow bandwidth enabled subthermionic switching}\label{RD_A}
	\vspace{-4.0mm}

    Here, we highlight the idea of reduced power dissipation or subthermionic switching in transistors by using low bandwidth materials. The $SS$ is defined as  
	\begin{align}\label{SS}
		SS \equiv \frac{\partial V_{\mathrm{GS}}}{{\partial log_{10} I_{\mathrm{DS}}}} = \underbrace{\frac{\partial V_{\mathrm{GS}}}{\partial \phi_b}}_{{m}} \times \underbrace{\frac{\partial \phi_b}{\partial log_{10} I_{\mathrm{DS}}}}_{{n}}~.
	\end{align}
	Here, $\phi_b$ is the barrier height from source-to-channel, and the first and second terms in Eq.~\eqref{SS} are the body factor ($m$) and the transport factor ($n$), respectively. %% is generally determined by the capacitive network of the transistor. The second term in (\ref{SS}) captures the conduction process and is called the transport factor ($n$). 
    To connect this to the electronic bandwidth, we note that 
		the electron density in a n-type 3-D semiconductor at a given energy is specified by the product of the density of states  and the Fermi function: $n(E)= D(E) \times f(E)$. Here, $f(E) = 1/[1+\exp(E/k_\mathrm{B} T)]$ is the Fermi function. For 3-D systems, we have $D_{3D}(E)\propto \sqrt{E}$. As a result, the carrier density in conventional bulk semiconductors like Si/Ge decays sub-exponentially with $E$, with an infinitely long thermal tail (see Figure \ref{FIG_1}). Similarly, the electron density in 2-D semiconductors is given by $D_{2D}(E) \propto E^0$ and we have $n(E) \propto f(E)$. This implies that the carrier density in 2-D systems also decays exponentially with $E$, and it also has an infinitely long thermal tail. Therefore, during the onset of conduction in conventional transistors comprised of 3D or 2D channel material, the current is determined by the the thermally activated carriers and we have $I_{\mathrm{DS}} \sim \exp(-\phi_t/\phi_b)$, with $\phi_t  \equiv k_B T/q$ being the temperature scale. This yields, $n \cong \ln(10) \phi_t $. %Here $k_B$ is the Boltzmann constant, $T$ is the temperature, $q$ is the elementary charge,  is the voltage equivalent of temperature. 
	In conventional transistors, the electrostatic coupling between gate and channel dictates that $m>1$. Thus, in conventional transistors, the $SS$ is limited to a value of $SS> 60$ mV/decade at room temperature (300 K), imposing a stiff limitation on their performance. It is typically referred to as the Boltzmann tyranny.
	%This natural limit on the $SS$ is known as the thermionic limit and is also referred to as Boltzmann's tyranny. Physically, it arises due to the long thermal tail of the charge carriers (see Figure~\ref{FIG_1}). 
    %However, in narrow bandwidth source materials, the above analysis for $I_{\rm DS}$ breaks down, and that allows for $n$ to have much smaller values and consequently we can have $SS < 60$ mV/decade. 
    %The presence of the thermally activated carriers due to the long thermal tail in both 3-D and 2-D devices, restricts their switching by 60 mV/decade at room temperature, imposing a stiff limitation on their performance. It is typically referred to as the Boltzmann tyranny.
	
	\begin{figure*}[!t]
		\includegraphics[width = 0.8\linewidth]{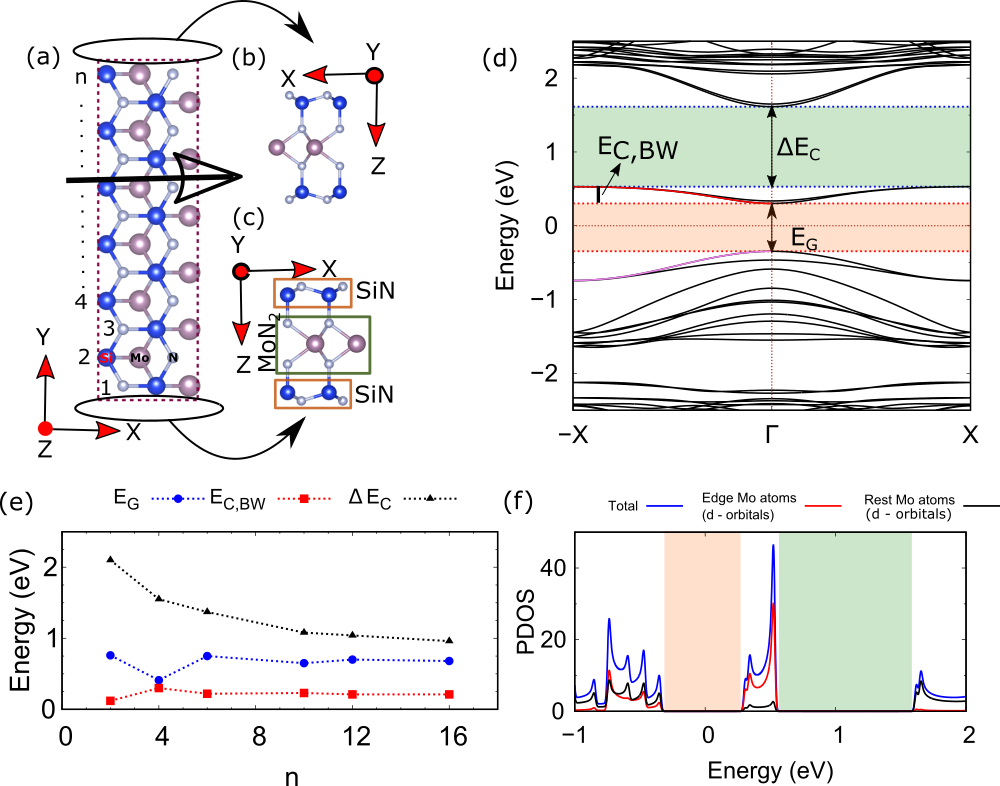}
		\caption{(a) Unit cell of a laterally confined (in the $y$-direction) monolayer of MoSi$_2$N$_4$. We refer to it as $n$-MoSi$_2$N$_4$, depending on the number ($n$) of bulk unit cells along the confinement direction. (b) and (c) show the zoomed view of the top and bottom edge of the structure, respectively. (d) The electronic bandstructure of  H-terminated 10-MoSi$_2$N$_4$. (e) Variation of the bandgap ($E _\mathrm{G}$), the bandwidth of the conduction band ($E_\mathrm{{C,BW}}$), and the gap of the first conduction band from the higher conduction band ($\Delta E_\mathrm{C}$) with the width of the nanoribbon in the confinement direction. (f) The atom and orbital resolved density of states of the edge and inner Mo atoms. Clearly, the narrow conduction band states arise from the edge Mo atoms. 
			\label{FIG_2}}
	\end{figure*}
	
	One possible solution to the problem of Boltzmann's tyranny arising from the long thermal tail, is to have a narrow bandwidth conduction band separated from the higher conduction band by a large energy gap. This cuts off the thermal tail above the potential barrier in the channel and suppresses the thermionic injection at a low gate-to-source bias ($V_\mathrm{{GS}}$) [see Figure \ref{FIG_1}]. In this scenario, direct source-to-drain tunneling (SDT) \cite{1175936} will leak a small amount of current in the OFF state ($I_\mathrm{{DS}}$), which strongly depends on the tunneling carrier's effective mass and width of the tunneling barrier (source-to-drain barrier). With the increase in $V_\mathrm{{GS}}$, the potential barrier in the channel decreases, and the current conduction mechanism changes from SDT to thermionic, resulting in a sudden increase in the drain current. This enables $n$ to have much smaller values and consequently we can have $SS < 60$ mV/decade, opening up possibilities for designing very power efficient transitors. 
	%the switching from OFF to ON state will be abrupt, and the device will facilitate sub-60 mV/decade switching. 
	This is the main idea, and highlight of this paper. In the rest of the paper, we demonstrate this idea of narrow band 2-D semiconductors overcoming Boltzmann's tyranny, using laterally confined MoSi$_2$N$_4$ as a channel material for realizing steep-slope and power-efficient logic electronic switches. To explicitly show this, we first establish the presence of a narrow bandwidth conduction band in laterally confined MoSi$_2$N$_4$ by calculating their electronic properties.

	\subsection{Electronic properties of H terminated MoSi$_2$N$_4$}\label{RD_B}
		\vspace{-4.0mm}
	We calculate the electronic properties of monolayer (2-D) MoSi$_2$N$_4$ for benchmarking our calculations [see Figure S1 and Table S1 in the Supporting Information (SI) 
	\footnote{The Supporting information has more discussions on the i) structural and electronic properties of the MoA$_2$Z$_4$ series of materials, ii) their passivated structures, iii) and the transport properties of transistors with nanoribbon of MoSi$_2$N$_4$ with different passivations.}]. 
	The results are found to be consistent with those reported in the literature \cite{Hong,wang2020structure}. We find the bandgap predicted by DFT with PBE functional (E$\rm_G$ = 1.86 eV)  to be very close to the experimental value (E$\rm _{G_{exp}}$ = 1.94 eV) \cite{Hong}. Orbital projected density of states (PDOS) reveals that in the vicinity of the conduction band minima (0.1 eV around CBM) $\sim$ 82 \% and $\sim$ 15 \% of the total DOS are contributed by the $d$-orbitals of molybdenum (Mo) and $p$-orbitals of nitrogen (N), respectively. Our charge distribution and orbital projected density of states (PDOS) analysis reveal that Mo-$d_{z^2}$ and N-$p_z$ orbitals hybridize to form the $\sigma$-bonds. As a result, bonding and anti-bonding orbitals form VBM and CBM, respectively [see Figure S2 of SI \cite{Note1} for further details]. 
	
	Starting from the optimized structure of 2-D monolayer, we construct $n$-MoSi$_2$N$_4$ nanoribbons (periodic along the $x$-direction) by stacking ($n/2$) number of unit cells along the $y$-direction [see Figure \ref{FIG_2}]. After creating the nanoribbon (say 10-MoSi$_2$N$_4$), the structure is relaxed once again before performing the electronic structure calculation. If dangling bonds of the edge atoms are left unpassivated, such a 10-MoSi$_2$N$_4$ nanoribbon is a semiconductor with a bandgap of E$\rm _G$ $\sim$ 0.9 eV [see Figure S3 of SI \cite{Note1} and related discussion].

	\begin{figure*}[!t]
		\includegraphics[width = .9\linewidth]{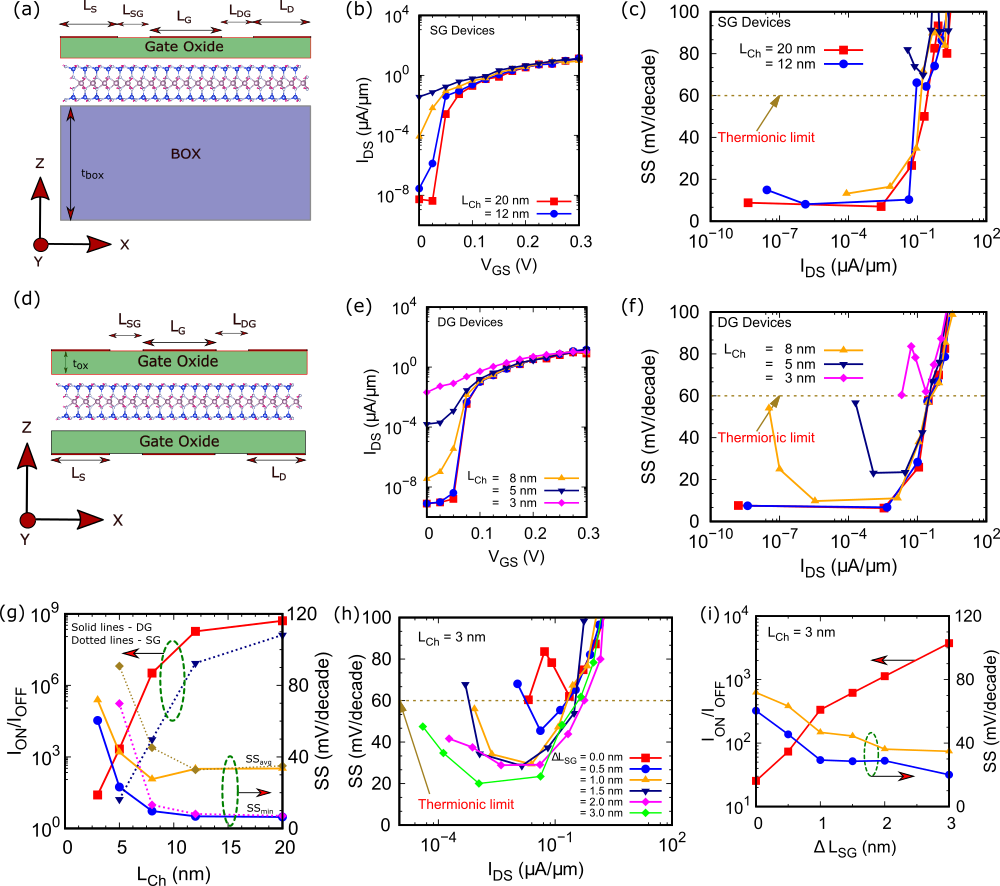}
		\caption{Schematic of a (a) single-gate (SG) device with  SiO$_2$ as the BOX and gate oxide with thickness $t_{\mathrm{box}}$ and $t_{\mathrm{ox}}$, respectively. (b) Transfer characteristics in semilogarithmic scale and (c) subthreshold slope vs drain current of SG devices for various channel lengths ($L_{\mathrm{Ch}}$). Schematic of (d) a double-gate (DG) devices and its (e) transfer characteristics in semilogarithmic scale and (f) subthreshold slope vs drain current for various $L_{\mathrm{Ch}}$. Both the SG and DG devices use H terminated 10-MoSi$_2$N$_4$ as the channel material. (g) $I_{\mathrm{ON}}/I_{\mathrm{OFF}}$ and SS vs $L_{\mathrm{Ch}}$ for SG and DG devices comprised of H terminated 10-MoSi$_2$N$_4$. (h) SS vs drain current and (i) $I_{\mathrm{ON}}/I_{\mathrm{OFF}}$ and SS vs $\Delta L_{\mathrm{SG}}$ for DG device for a H terminated 10-MoSi$_2$N$_4$ with $L_{\mathrm{Ch}}$ = 3 nm. The devices with $SS < 60$ mV/decade indicate highly power efficient transistors. 
			\label{FIG_3}}
	\end{figure*}
	
	To eliminate the effect of the dangling bonds, we use hydrogen (H) atoms to terminate the edges. Hydrogen is commonly used for edge termination in experiments \cite{NL_H_GNR,NL_H_GNR_1,Morita}, as the process can be easily controlled by the pressure and temperature of the H$_2$ gas. To keep the coordination number unchanged, each of the edge Mo, N, and Si atoms is terminated by two H atoms, one H atom, and one H atom, respectively. The electronic band structure of the optimized structure is shown in Figure~\ref{FIG_2} (d), along the high symmetry path (-X - $\Gamma$ - X). The two lowest energy conduction bands have narrow bandwidth (E$\rm _{C,BW}$), and they are separated from nearby bands by a relatively large gap of $\Delta$E$\rm_C$. The variation of E$\rm _G$, E$\rm _{C,BW}$, and $\Delta$E$\rm_C$ with $n$ are plotted in Figure \ref{FIG_2} (e) for H-terminated $n$-MoSi$_2$N$_4$. As $n$ increases, E$\rm _G$, E$\rm _{C,BW}$, and $\Delta$E$\rm_C$ saturate to $\sim$ 0.7 eV, $\sim$ 0.2 eV, and $\sim$ 1.0 eV, respectively. These localized conduction bands, in H-terminated 10-MoSi$_2$N$_4$, are predominated by $d$-orbitals of edge Mo atoms, as shown in Figure~\ref{FIG_2} (f). See Figure~S4 of SI \cite{Note1} for more details. Such a unique combination of electronic states can potentially achieve the target of sub-60 mV/decade switching in transistors by cutting off the thermionic current in the OFF-state.

	\subsection{Transistors based on H-terminated MoSi$_2$N$_4$}\label{RD_C}
		\vspace{-4.0mm}
	To evaluate the the transport properties and sub-60 mV/decade switching, we perform first-principles-based nanoscale device simulation of single-gate (SG) and double-gate (DG) devices composed of H-terminated 10-MoSi$_2$N$_4$. The schematics of devices are shown in Figure~\ref{FIG_3} (a) and (d), in which the main geometrical parameters are the gate length ($L_{\mathrm{G}}$), source/drain length ($L_{\mathrm{S/D}}$), source/drain-to-gate length ($L_{\mathrm{SG/DG}}$), and oxide thickness ($t_{\mathrm{ox}}$). SiO$_2$ is used as both gate oxide [$t_{\mathrm{ox}}$ = 0.60 nm] and box oxide [$t_{\mathrm{box}}$ = 10 nm]. The source/drain is electrostatically doped using source/drain gate with $L_{\mathrm{S/D}}$ = 9.80 nm, and $L_{\mathrm{SG/DG}}$ = 0.20 nm. The power supply voltage ($V_{\mathrm{DD}}$) is 0.1 V.
	The devices' characteristics are shown by considering 200 parallelly connected chains of devices per $\mu$m along the z-direction \cite{qiu2018dirac}.
	
	\begin{figure}[!t]
		\includegraphics[width = 0.99\linewidth]{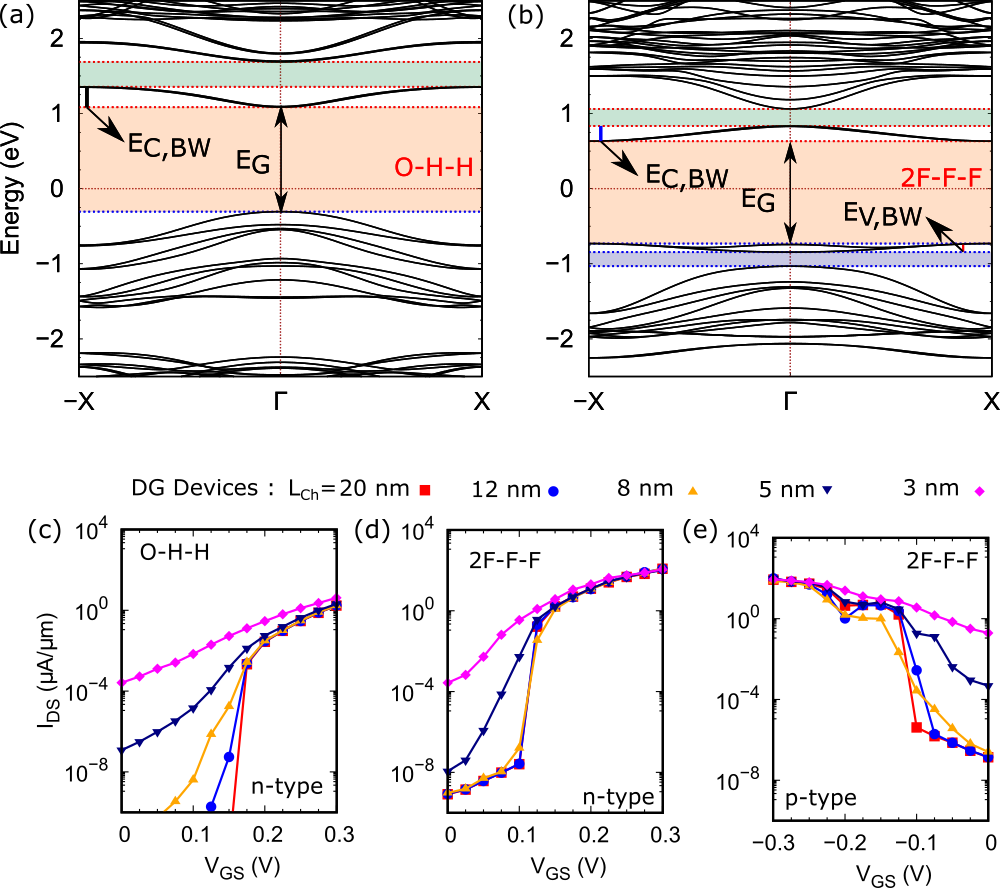}
		\caption{Narrow bandwidth electronic band structure of (a) O-H-H and (b) 2F-F-F terminated 10-MoSi$_2$N$_4$. In O-H-H terminated structure, the two lowest energy conduction bands have a narrow bandwidth. %in energy with width $\rm E_{C,BW}$. 
			Interestingly, the two highest energy valance bands are also localized in energy with width  $\rm E_{V,BW}$ in addition to the bottom two energy localized conduction bands. The transport properties of DG devices include (c) O-H-H and (d, e) 2F-F-F as a channel material with $L_{\mathrm{Ch}}$ = 20 nm, 12 nm, 8 nm, 5 nm, and 3 nm.
			\label{FIG_4}}
	\end{figure}
	
	The variation of the drain current with gate potential (transfer characteristics) and that of the $SS$ with drain current for both SG and DG devices for various channel lengths $L_{\mathrm{Ch}}$, are shown in Figure~\ref{FIG_3} (b, e) and in (c, f), respectively. We find that over more than three decades of drain current in SG devices, the drain current exhibits subthermionic nature, with an average subthreshold slope of less than 46 mV/decade for $L_\mathrm {Ch}$ $\le$ 8 nm. They also show a good ON-OFF current ratio ($> 10^3$) with the minimum point subthreshold slope less than 25 mV/decade down to 8 nm channel length. The performance parameters of DG devices with $L_\mathrm{Ch}$ $\le$ 5 nm are the same as those of SG devices with $L_\mathrm{Ch}$ $\le$ 8 nm.
	Further scaling causes excessive source-to-drain leakage and the loss of subthermionic switching in SG and DG devices with $L_\mathrm{Ch} \le 5$ nm and $L_\mathrm{Ch} \le 3$ nm, as highlighted in Figure~\ref{FIG_3} (g).
	Direct source-to-drain tunneling limits how steeply these devices can turn on. This tunneling is proportional to $\exp(-\sqrt m_{\rm eff} \times l_t)$, where $m_{\rm eff}$ is the effective mass along the transport direction and $l_t$ is the tunneling length. Hence, the rate of increase of the drain current when going from the OFF state to the ON state 
	degrades with the reduction of the channel length of the device. We find that the rate of degradation is lower in DG devices compared to SG devices due to better electrostatics in DG devices.

	To counter the impact of increased source-to-drain tunneling, we increase the channel length of the double gate device with a gate length of 3 nm by adding $\Delta L_\mathrm{SG}$ to $L_\mathrm{SG/DG}$.
	The resulting device structure effectively increases the source-drain barrier width and height, thereby reducing the source-to-drain leakage and subthreshold leakage. However, there is a trade-off between $\Delta L_\mathrm{SG}$ and ON-state performance due to reduced gate control over the extended channel \cite{1386595,Keshari_TED_PdSe2}. 
	The simulated transport characteristics for various $\Delta L_\mathrm{SG}$ values up to 3 nm [0, 0.5, 1.0, 1.5, 2.0, 3.0 nm] are shown in Figure S10 of the SI \cite{Note1}.
	We find that with increase in $\Delta L_{\mathrm{SG}}$, the source-to-drain leakage decreases. The device exhibits subthermionic behavior with a $SS_{\mathrm{min}}$ $<$ 30 mV/decade, and $SS_{\mathrm{avg}}$ $<$ 37 mV/decade over more than three decades of drain current at $\Delta L_\mathrm{SG}$ $\ge$ 2.0 nm, with an ON-OFF ratio more than $10^3$, as shown in Figure~\ref{FIG_3} (h) and (i).

	\subsection{Edge Functionalization of MoSi$_2$N$_4$}\label{RD_D}
		\vspace{-4.0mm}
	Having demonstrated efficient subthreshold switching below 60 mV/decade in laterally confined MoSi$_2$N$_4$ devices with hydrogen passivation, a natural question to ask is what happens on passivating with other atoms \cite{Lopez_JPCC,Ren2018_SR,Zhu_JPCC,Lopez_Nano}. To address this, we subject 10-MoSi$_2$N$_4$ to different edge functionalization, by passivating the dangling bonds with H, F, and O, which are commonly used for edge termination \cite{Lopez_JPCC,Sodi_PRB,D0TC01764G,Mirco_Nano,Lopez_Nano,Zhu_JPCC,Ren2018_SR}. In exploring edge functionalization, we keep the coordination number of edge atoms to be the same as in the bulk 2D sample. We investigate several possible combinations of edge functionalization, which are listed in Table~\ref{table:nonlin}.
	
	\begin{table}[!b]
		\caption{Different edge functionalization  schemes for MoSi$_2$N$_4$.} % title of Table
		\centering % used for centering table
		\begin{tabular}{c  c  c c %>{\columncolor{lime}}
				c} % centered columns (4 columns)
			\hline\hline %inserts double horizontal lines
			%\rowcolor{lightgray}
			S. No.	& Edge Atoms Termination Scheme & & &Notation \\ [0.5ex] % inserts table
			%heading
			\hline 
			\hline % inserts single horizontal line
			1 & (Mo, Si, N) $\to$ (2H, H, H) & & &2H-H-H\\
			2 & (Mo, Si, N) $\to$ (2F, F, F) & & &2F-F-F\\
			3&(Mo, Si, N) $\to$  (2F, H, H)& & & 2F-H-H\\ 
			4&(Mo, Si, N) $\to$  (2H, F, F)& & &2H-F-F\\
			5&(Mo, Si, N) $\to$  (O, H, H) & & &O-H-H\\
			6&(Mo, Si, N) $\to$  (O, F, F) & & &O-F-F\\
			7&(Mo-1, Mo-2, Si, N) $\to$ (H, F, H, H) & & & HF-H-H\\
			8&(Mo, Si, N) $\to$  (N, H, H) & & &N-H-H\\ [1ex] % [1ex] adds vertical space
			\hline 
			\hline %inserts single line
		\end{tabular}
		\label{table:nonlin} % is used to refer this table in the text
	\end{table}
	
	Among all the edge functionalized MoSi$_2$N$_4$ nanoribbons mentioned above, two of the band structures are shown in Figure~\ref{FIG_4}.  In case of O-H-H terminated 10-MoSi$_2$N$_4$ shown in Figure~\ref{FIG_4} (a),  E$\rm _G$ ($\sim$ 1.40 eV) is larger than the bandgap of 2H-H-H terminated 10-MoSi$_2$N$_4$. The two lowest energy conduction bands have a bandwidth of E$\rm _{C,BW}$ $\sim$ 0.27 eV, and are separated from the next conduction band by $\sim$ 0.33 eV. Our Bader charge analysis \cite{HENKELMAN2006354,Bader1} shows that the charge on each Mo atom is reduced by 0.31$e$ in O-H-H terminated 10-MoSi$_2$N$_4$ compared to that in 2H-H-H terminated 10-MoSi$_2$N$_4$. As a result, the percentage contribution of the $d$-orbital of the edge Mo atoms is less (CBM 40\% and VBM 15\%) in O-H-H terminated 10-MoSi$_2$N$_4$, than that in 2H-H-H terminated 10-MoSi$_2$N$_4$ (CBM 67\% and VBM 24 \%). For 2F-F-F terminated 10-MoSi$_2$N$_4$ [see Figure~\ref{FIG_4} (b)], the bandgap ($\sim$ 1.36 eV) is larger than the bandgap of purely 2H-H-H terminated 10-MoSi$_2$N$_4$. The two lowest energy conduction bands have a bandwidth of E$\rm _{C,BW}$ $\sim$ 0.20 eV, and these are separated from the next conduction band by $\sim$ 0.23 eV. Interestingly, we find that 2F-F-F terminated nanoribbon shows high energy filtering capability for holes also, having E$\rm _{V,BW}$ $\sim$ 0.12 eV and separated from the nearest valence band by $\sim$ 0.19 eV [see Figure \ref{FIG_4} (b)]. Hence, it can be used to realize both n- and p-type steep-slope devices. 
	
	\begin{figure}[t!]
		\includegraphics[width = 0.99\linewidth]{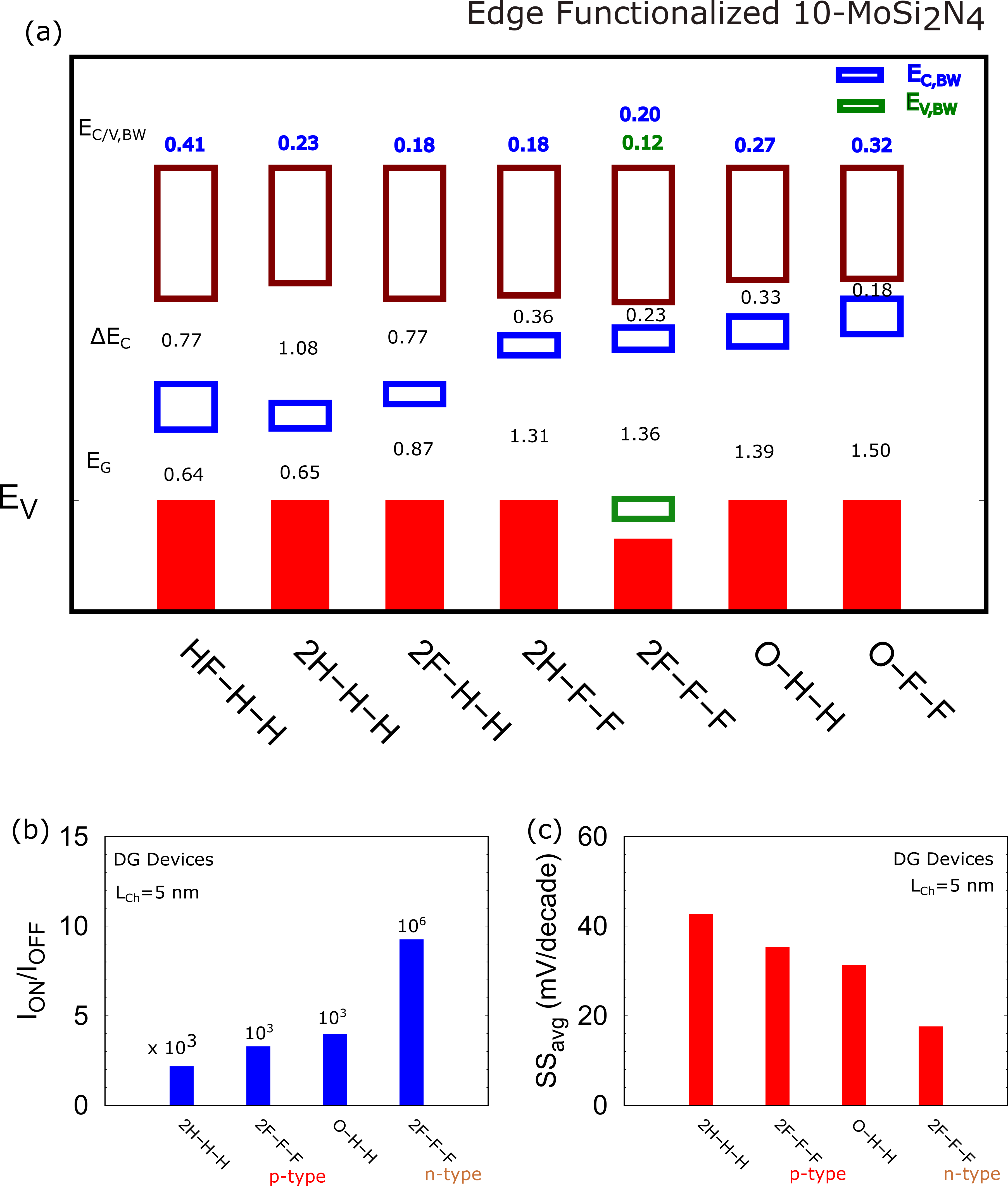}
		\caption{(a) Pictorial representation of the unique electronic properties of 10-MoSi$_2$N$_4$ with different terminations, such as HF-H-H, 2F-H-H, 2H-F-F, O-H-H, and O-F-F. All the studied structures have a narrow lowest energy conduction band which is separated from the higher-energy conduction band. 2F-F-F terminated 10-MoSi$_2$N$_4$ also has an isolated narrow valance band which can enable subthermionic switching in a p-type device. (b) Excellent ON/OFF ratio and (c) very low average subthreshold slope of DG devices based on 2H-H-H, O-H-H, and 2F-F-F terminated 10-MoSi$_2$N$_4$ as a channel material with  $L_{\mathrm{Ch}}$ = 5 nm. While all of these offer extreme power efficiency in terms of transistor performance, the performance parameters for 2F-F-F terminated 10-MoSi$_2$N$_4$ are amongst the best reported values for any 2D material. 
			\label{fig_5}}
	\end{figure}
	
	The most remarkable outcome of our detailed analysis of the electronic properties of edge functionalized MoSi$_2$N$_4$ nanoribbon is its robust band structure, which has the same qualitative features irrespective of the choice of edge functionalization. Band diagrams of several possible combinations of edge functionalization are compared in Figure~\ref{fig_5} (a), illustrating a wide range of values for E$\rm _G$ (0.64 eV - 1.50 eV) and $\Delta$E$\rm_C$ (0.18 eV - 1.08 eV), while E$\rm _{C,BW}$ value lies within a small range of 0.42 eV - 0.18 eV. The corresponding band structures are shown in Figure~S5 of SI \cite{Note1}. 
	Most of these structures have a narrow bandwidth conduction band which limits the long thermal tail of carriers and thermionic injection in the OFF state. This makes them suitable for low-power applications. Additionally, edge functionalization engineering offers a route to control their desired electronic properties. 
	Amongst these, the 2F-F-F terminated nanoribbon is the only one having both narrow conduction and narrow valance bands. In addition, we find that the N-H-H terminated nanoribbon is metallic [see Figure~S5 of SI \cite{Note1}]. 
	In fact, such `cold' metals can be used as a cold source contact in FETs, which is another way to achieve sub-60 mV/decade switching and overcome Boltzmann's tyranny \cite{PhysRevApplied.13.064037}.
	
	\subsection{Transitors with functionalized MoSi$_2$N$_4$ and other materials in the series}\label{RD_E}
		\vspace{-4.0mm}
	The transfer characteristics of DG devices, with O-H-H terminated n-type, 2F-F-F terminated n-type, and 2F-F-F terminated p-type 10-MoSi$_2$N$_4$ as channel material is shown in Figure~\ref{FIG_4} (c-e). All the devices show subthermionic nature in drain current down to 5 nm channel length. More specifically, the drain current show subthermionic nature over more than five decades, eight decades, and three decades of drain current for devices with $L_{\mathrm{Ch}}$ $\le$ 5 nm based on n-type devices with O-H-H, and 2F-F-F termination, and p-type device with 2F-F-F termination, respectively. 
	
	% 	\begin{figure}[!t]
		% 		\includegraphics[width = 1.0\linewidth]{Figures/FIG_5.png}
		% 		\caption{(a) Excellent ON/OFF ratio and (b) very low averge subthreshold slope of DG devices based on 2H-H-H, O-H-H, and 2F-F-F terminated 10-MoSi$_2$N$_4$ as a channel material with  $L_{\mathrm{Ch}}$ = 5 nm. While all of these offer extreme power efficiency in terms of transistor performance, the performance parameters for 2F-F-F terminated 10-MoSi$_2$N$_4$ are amongst the best reported values for any 2D material.  
			% 		\label{FIG_6}}
		% 	\end{figure}
	% 	%
	The best performance of DG devices with $L_{\mathrm{Ch}}$ = 5 nm is shown in Figure \ref{fig_5} (b) and (c).
	All the devices show ON/OFF ratio of more than $10^3$ and an average subthreshold slope of less than 45 mV/decade. However, the best performing device is n-type comprised of 2F-F-F terminated 10-MoSi$_2$N$_4$, owing to flat conduction band minima and hence low SDT leakage, with $I_{\mathrm{ON}}/I_{\mathrm{OFF}}$ $>$ $10^6$ and  $SS_{\mathrm{avg}}$ $<$ 20 mV/decade. See Sec.~E of the SI \cite{Note1} for a more detailed analysis. The upper limit of devices' performance is assessed as the ballistic nature of transport is considered. However, the presence of electron-phonon interaction could degrade the device's performance, but the subthermionic nature of devices could be maintained \cite{PRR_D, logoteta2021intrinsic}.
	
\begin{figure}[!t]
		\includegraphics[width = 0.99\linewidth]{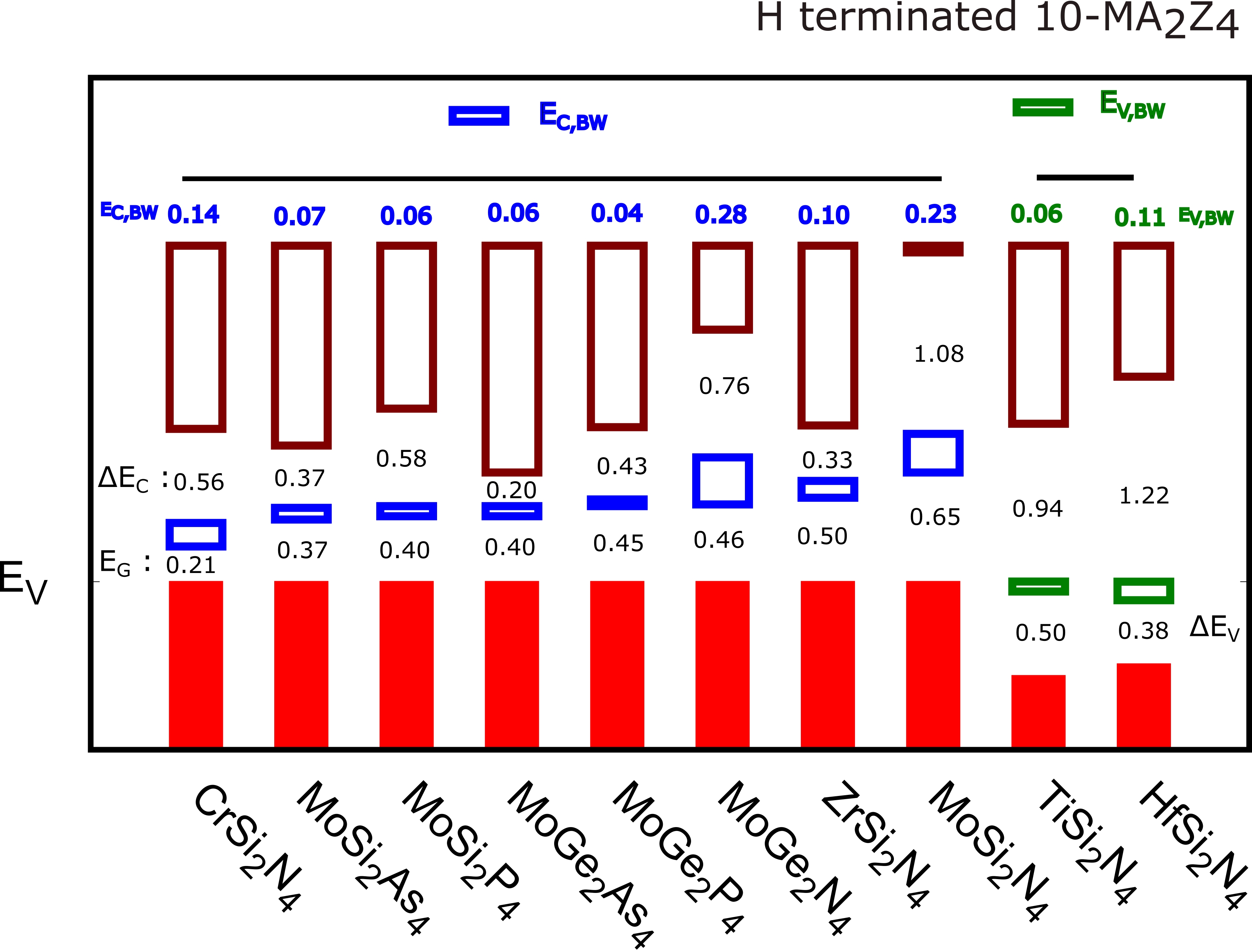}
		\caption{Pictorial representation of unique electronic properties different materials of the MA$_2$Z$_4$ series, with H termination. Going beyond MoSi$_2$N$_4$, several other materials in the same series also have similar bandstructure features as laterally confined MoSi$_2$N$_4$, which enable subthermionic switching for power efficient transistors. 
			\label{fig_7}}
	\end{figure}

	Having demonstrated excellent low-power transistor characteristics for laterally confined and passivated MoSi$_2$N$_4$, we now explore other materials in the same series of compounds for ultra-low-power devices. Specifically, we extend the search in MA$_2$Z$_4$ the family, where M=Mo, Cr, Zr, Ti, Hf; A=Si, Ge; and Z=N, P, As. The calculated band diagrams for 2H-H-H terminated 10-MA$_2$Z$_4$ nanoribbon is shown in Figure~\ref{fig_7}. We find that there are several candidates in this family, having narrow bandwidth first valence or conduction bands, separated from nearby bands by a sizeable gap. This opens up the field for further exploration of MA$_2$Z$_4$ series of materials, for potential application in ultra-low power transistor applications as channel materials. \\
	
	\section{Conclusion}\label{Conclusion}
	\vspace{-4.0mm}
	One of the major bottlenecks for making ultra-power-efficient transistors is decreasing the applied gate voltage for a 10-fold increase in the source-to-drain current during the conduction process. The bottleneck is controlled by the long thermal tail of the charge carriers, which leads to a large thermionic current in the OFF state. We propose a solution to this problem by using narrow bandwidth semiconductors as channel material. The narrow bandwidth of the conduction/valance band naturally restricts the long thermal tail of the charge carriers, reduces the thermionic current in the OFF state, and solves the problem of Boltzmann's tyranny.
	
	We demonstrate this using the example of the laterally confined monolayer MoSi$_2$N$_4$. We show that passivated MoSi$_2$N$_4$ nano-ribbons have a narrow bandwidth conduction band, which is separated from nearby bands by a large bandgap. By combining first-principle calculations and nano-scale device simulations, we evaluate the performance of several transistors using differently terminated MoSi$_2$N$_4$ nano-ribbons as channel material. We find that devices with several differently terminated MoSi$_2$N$_4$ as channel material show a considerable performance improvement with a subthreshold slope much smaller than 60 mV/decade. Some of them can be used to realize both n-and p-type power-efficient transistors. We show that these unique electronic properties of narrow bandwidth are also shared by several other members of the MSi$_2$N$_4$ family of materials. Our work offers a new direction for engineering and exploring narrow bandwidth materials for enabling steep-slope transistors and ultra power-efficient computing.\\ 
	
	\appendix
	\section{Methods} 
		\vspace{-4.0mm}
	\subsection{Density Functional Theory Calculations} 
		\vspace{-4.0mm}
	All the DFT-based first-principles calculations are performed using the projector augmented wave (PAW) potentials, as implemented in the Quantum ESPRESSO code suite \cite{QE_1,QE_2}. A Perdew-Burke-Ernzerhof (PBE) \cite{PBE} based generalized gradient approximation (GGA) is used to include the exchange and correlation effects. The kinetic energy cut-off of the wave function and the charge density is set to 60 Ry and 600 Ry, respectively. The charge transfer is analyzed using the Bader partitioning scheme, in which the electron density is divided at zero flux surfaces \cite{HENKELMAN2006354,Bader1}. A $14 \times 14 \times 1$ Monkhorst-Pack (MP) $k-$mesh is used for Brillouin zone integration. For lateral confinement, a thick vacuum layer ($\sim$ 25 \AA) is considered in $y$ and $z$ direction.
	
	\subsection{Device Simulations}
		\vspace{-4.0mm}
	The electronic band structure of the system is represented by a set of Wannier functions to upscale the system for device simulation via the Wannier90 code suite \cite{W90}. The overlap matrix between the adjacent $k$ points and the projection matrix is used as the input to calculate the Wannier functions. The resulting transformed Hamiltonian matrix in the basis of MLWFs is sparse in nature.
	We calculate the transmission coefficients [$T(E)$] and potentials at each bias ($V_\mathrm{GS}$, $V_\mathrm{DS}$), by solving the Poisson and Schr$ \rm \ddot{o}$dinger equations self-consistently in the NEGF formalism \cite{datta2005quantum,NanoTCAD}. 
	The Landauer-B$\rm \ddot{u}$ttiker method is used to estimate the drain current, which can be expressed as
	\begin{equation}\label{Landauer_formulation}
		\nonumber
		I_{DS} = \frac{2e}{h} \int_{-\infty}^{\infty} T(E,V_{GS},V_{DS}) [f(E-\mu_S)-f(E-\mu_D)] dE~.
	\end{equation}
	Here, $e$ is the elementary charge, $h$ is the Planck constant and $V_\mathrm{GS}$ ($V_\mathrm{DS}$) is the gate-source (drain-source) voltage. $T(E,V_{GS},V_{DS})$ is the transmission coefficient at a given energy $E$ and a bias $(V_{GS},V_{DS})$. $\mu_{S/D}$ and $f(E-\mu_{S/D})$ are the fermi energy level and the Fermi-Dirac distribution function at source/drain, respectively. The tunneling probability,  $T(E,V_{GS},V_{DS})$ is calculated using
	\begin{align}\label{Transmission}
		T(E,V_{GS},V_{DS}) = {\rm Trace}[\Gamma_S G^R \Gamma_D G^A]~.
	\end{align}
	Here, $G^{R/A}$ is the retarded/advance Green functions, $\Gamma_{S/D}$ is the broadening from source/drain contacts and $E$ is the energy. Additionally, we have 
	\begin{subequations}
		\begin{align}
			\Gamma_{S/D}(E,V_{GS},V_{DS}) = i[\Sigma_{S/D}-\Sigma^\dagger_{S/D}]~,\\
			G^R(E,V_{GS},V_{DS}) = [EI-H-\Sigma]^{-1}~,\\
			G^A(E,V_{GS},V_{DS}) =[G^R(E,V_{GS},V_{DS})]^\dagger~.
		\end{align}
	\end{subequations}
	Here, $I$ is the an identity matrix,  $\Sigma_{S/D}$ is source/drain contact self-energy, $\Sigma$ = $\Sigma_{S}$+$\Sigma_{D}$, and $H$ is the Hamiltonian matrix representing the channel. \\

	\section*{Acknowledgements}
		\vspace{-4.0mm}
	This work was supported in part by the Humboldt Foundation and in part by the Swarnajayanti Fellowship of the Department of Science and Technology (DST), Government of India under Grant DST/SJF/ETA-02/2017-18. The authors thank the CC-IITK for providing the HPC facility. We acknowledge the Science and Engineering Research Board (SERB, for projects MTR/2019/001520), and the Department of Science and Technology (DST, for project DST/NM/TUE/QM-6/2019(G)-IIT Kanpur) of the Government of India for financial support.
	
	\bibliography{References}% Produces the bibliography via BibTeX.

\providecommand{\noopsort}[1]{}\providecommand{\singleletter}[1]{#1}%
\begin{thebibliography}{68}%
\makeatletter
\providecommand \@ifxundefined [1]{%
 \@ifx{#1\undefined}
}%
\providecommand \@ifnum [1]{%
 \ifnum #1\expandafter \@firstoftwo
 \else \expandafter \@secondoftwo
 \fi
}%
\providecommand \@ifx [1]{%
 \ifx #1\expandafter \@firstoftwo
 \else \expandafter \@secondoftwo
 \fi
}%
\providecommand \natexlab [1]{#1}%
\providecommand \enquote  [1]{``#1''}%
\providecommand \bibnamefont  [1]{#1}%
\providecommand \bibfnamefont [1]{#1}%
\providecommand \citenamefont [1]{#1}%
\providecommand \href@noop [0]{\@secondoftwo}%
\providecommand \href [0]{\begingroup \@sanitize@url \@href}%
\providecommand \@href[1]{\@@startlink{#1}\@@href}%
\providecommand \@@href[1]{\endgroup#1\@@endlink}%
\providecommand \@sanitize@url [0]{\catcode `\\12\catcode `\$12\catcode
  `\&12\catcode `\#12\catcode `\^12\catcode `\_12\catcode `\%12\relax}%
\providecommand \@@startlink[1]{}%
\providecommand \@@endlink[0]{}%
\providecommand \url  [0]{\begingroup\@sanitize@url \@url }%
\providecommand \@url [1]{\endgroup\@href {#1}{\urlprefix }}%
\providecommand \urlprefix  [0]{URL }%
\providecommand \Eprint [0]{\href }%
\providecommand \doibase [0]{http://dx.doi.org/}%
\providecommand \selectlanguage [0]{\@gobble}%
\providecommand \bibinfo  [0]{\@secondoftwo}%
\providecommand \bibfield  [0]{\@secondoftwo}%
\providecommand \translation [1]{[#1]}%
\providecommand \BibitemOpen [0]{}%
\providecommand \bibitemStop [0]{}%
\providecommand \bibitemNoStop [0]{.\EOS\space}%
\providecommand \EOS [0]{\spacefactor3000\relax}%
\providecommand \BibitemShut  [1]{\csname bibitem#1\endcsname}%
\let\auto@bib@innerbib\@empty
%</preamble>
\bibitem [{\citenamefont {Ionescu}\ and\ \citenamefont {Riel}(2011)}]{Adrian}%
  \BibitemOpen
  \bibfield  {author} {\bibinfo {author} {\bibfnamefont {Adrian~M.}\
  \bibnamefont {Ionescu}}\ and\ \bibinfo {author} {\bibfnamefont {Heike}\
  \bibnamefont {Riel}},\ }\bibfield  {title} {\enquote {\bibinfo {title}
  {Tunnel field-effect transistors as energy-efficient electronic switches},}\
  }\href {\doibase 10.1038/nature10679} {\bibfield  {journal} {\bibinfo
  {journal} {Nature}\ }\textbf {\bibinfo {volume} {479}},\ \bibinfo {pages}
  {329--337} (\bibinfo {year} {2011})}\BibitemShut {NoStop}%
\bibitem [{\citenamefont {Zhang}\ \emph {et~al.}(2006)\citenamefont {Zhang},
  \citenamefont {Zhao},\ and\ \citenamefont {Seabaugh}}]{Qin_TFET}%
  \BibitemOpen
  \bibfield  {author} {\bibinfo {author} {\bibfnamefont {Qin}\ \bibnamefont
  {Zhang}}, \bibinfo {author} {\bibfnamefont {Wei}\ \bibnamefont {Zhao}}, \
  and\ \bibinfo {author} {\bibfnamefont {A.}~\bibnamefont {Seabaugh}},\
  }\bibfield  {title} {\enquote {\bibinfo {title} {Low-subthreshold-swing
  tunnel transistors},}\ }\href {\doibase 10.1109/LED.2006.871855} {\bibfield
  {journal} {\bibinfo  {journal} {IEEE Electron Device Letters}\ }\textbf
  {\bibinfo {volume} {27}},\ \bibinfo {pages} {297--300} (\bibinfo {year}
  {2006})}\BibitemShut {NoStop}%
\bibitem [{\citenamefont {Nagavarapu}\ \emph {et~al.}(2008)\citenamefont
  {Nagavarapu}, \citenamefont {Jhaveri},\ and\ \citenamefont {Woo}}]{4457828}%
  \BibitemOpen
  \bibfield  {author} {\bibinfo {author} {\bibfnamefont {Venkatagirish}\
  \bibnamefont {Nagavarapu}}, \bibinfo {author} {\bibfnamefont {Ritesh}\
  \bibnamefont {Jhaveri}}, \ and\ \bibinfo {author} {\bibfnamefont {Jason
  C.~S.}\ \bibnamefont {Woo}},\ }\bibfield  {title} {\enquote {\bibinfo {title}
  {The tunnel source (pnpn) n-mosfet: A novel high performance transistor},}\
  }\href {\doibase 10.1109/TED.2008.916711} {\bibfield  {journal} {\bibinfo
  {journal} {IEEE Transactions on Electron Devices}\ }\textbf {\bibinfo
  {volume} {55}},\ \bibinfo {pages} {1013--1019} (\bibinfo {year}
  {2008})}\BibitemShut {NoStop}%
\bibitem [{\citenamefont {Appenzeller}\ \emph {et~al.}(2004)\citenamefont
  {Appenzeller}, \citenamefont {Lin}, \citenamefont {Knoch},\ and\
  \citenamefont {Avouris}}]{Knoch_PRL}%
  \BibitemOpen
  \bibfield  {author} {\bibinfo {author} {\bibfnamefont {J.}~\bibnamefont
  {Appenzeller}}, \bibinfo {author} {\bibfnamefont {Y.-M.}\ \bibnamefont
  {Lin}}, \bibinfo {author} {\bibfnamefont {J.}~\bibnamefont {Knoch}}, \ and\
  \bibinfo {author} {\bibfnamefont {Ph.}\ \bibnamefont {Avouris}},\ }\bibfield
  {title} {\enquote {\bibinfo {title} {Band-to-band tunneling in carbon
  nanotube field-effect transistors},}\ }\href {\doibase
  10.1103/PhysRevLett.93.196805} {\bibfield  {journal} {\bibinfo  {journal}
  {Phys. Rev. Lett.}\ }\textbf {\bibinfo {volume} {93}},\ \bibinfo {pages}
  {196805} (\bibinfo {year} {2004})}\BibitemShut {NoStop}%
\bibitem [{\citenamefont {Salahuddin}\ and\ \citenamefont
  {Datta}(2008)}]{Sayeef}%
  \BibitemOpen
  \bibfield  {author} {\bibinfo {author} {\bibfnamefont {Sayeef}\ \bibnamefont
  {Salahuddin}}\ and\ \bibinfo {author} {\bibfnamefont {Supriyo}\ \bibnamefont
  {Datta}},\ }\bibfield  {title} {\enquote {\bibinfo {title} {Use of negative
  capacitance to provide voltage amplification for low power nanoscale
  devices},}\ }\href {\doibase 10.1021/nl071804g} {\bibfield  {journal}
  {\bibinfo  {journal} {Nano Letters}\ }\textbf {\bibinfo {volume} {8}},\
  \bibinfo {pages} {405--410} (\bibinfo {year} {2008})}\BibitemShut {NoStop}%
\bibitem [{\citenamefont {Salvatore}\ \emph {et~al.}(2008)\citenamefont
  {Salvatore}, \citenamefont {Bouvet},\ and\ \citenamefont
  {Ionescu}}]{adrian_ncfet}%
  \BibitemOpen
  \bibfield  {author} {\bibinfo {author} {\bibfnamefont {Giovanni~A.}\
  \bibnamefont {Salvatore}}, \bibinfo {author} {\bibfnamefont {Didier}\
  \bibnamefont {Bouvet}}, \ and\ \bibinfo {author} {\bibfnamefont
  {Adrian~Mihai}\ \bibnamefont {Ionescu}},\ }\bibfield  {title} {\enquote
  {\bibinfo {title} {Demonstration of subthrehold swing smaller than
  60mv/decade in fe-fet with p(vdf-trfe)/sio2 gate stack},}\ }\href {\doibase
  10.1109/IEDM.2008.4796642} {\bibfield  {journal} {\bibinfo  {journal} {2008
  IEEE International Electron Devices Meeting}\ ,\ \bibinfo {pages} {1--4}}
  (\bibinfo {year} {2008})}\BibitemShut {NoStop}%
\bibitem [{\citenamefont {Rusu}\ \emph {et~al.}(2010)\citenamefont {Rusu},
  \citenamefont {Salvatore}, \citenamefont {Jiménez},\ and\ \citenamefont
  {Ionescu}}]{Rusu}%
  \BibitemOpen
  \bibfield  {author} {\bibinfo {author} {\bibfnamefont {Alexandru}\
  \bibnamefont {Rusu}}, \bibinfo {author} {\bibfnamefont {Giovanni~A.}\
  \bibnamefont {Salvatore}}, \bibinfo {author} {\bibfnamefont {David}\
  \bibnamefont {Jiménez}}, \ and\ \bibinfo {author} {\bibfnamefont
  {Adrian~M.}\ \bibnamefont {Ionescu}},\ }\bibfield  {title} {\enquote
  {\bibinfo {title} {Metal-ferroelectric-meta-oxide-semiconductor field effect
  transistor with sub-60mv/decade subthreshold swing and internal voltage
  amplification},}\ }in\ \href {\doibase 10.1109/IEDM.2010.5703374} {\emph
  {\bibinfo {booktitle} {2010 International Electron Devices Meeting}}}\
  (\bibinfo {year} {2010})\ pp.\ \bibinfo {pages} {16.3.1--16.3.4}\BibitemShut
  {NoStop}%
\bibitem [{\citenamefont {Si}\ \emph {et~al.}(2018)\citenamefont {Si},
  \citenamefont {Su}, \citenamefont {Jiang}, \citenamefont {Conrad},
  \citenamefont {Zhou}, \citenamefont {Maize}, \citenamefont {Qiu},
  \citenamefont {Wu}, \citenamefont {Shakouri}, \citenamefont {Alam},\ and\
  \citenamefont {Ye}}]{Si2018}%
  \BibitemOpen
  \bibfield  {author} {\bibinfo {author} {\bibfnamefont {Mengwei}\ \bibnamefont
  {Si}}, \bibinfo {author} {\bibfnamefont {Chun-Jung}\ \bibnamefont {Su}},
  \bibinfo {author} {\bibfnamefont {Chunsheng}\ \bibnamefont {Jiang}}, \bibinfo
  {author} {\bibfnamefont {Nathan~J.}\ \bibnamefont {Conrad}}, \bibinfo
  {author} {\bibfnamefont {Hong}\ \bibnamefont {Zhou}}, \bibinfo {author}
  {\bibfnamefont {Kerry~D.}\ \bibnamefont {Maize}}, \bibinfo {author}
  {\bibfnamefont {Gang}\ \bibnamefont {Qiu}}, \bibinfo {author} {\bibfnamefont
  {Chien-Ting}\ \bibnamefont {Wu}}, \bibinfo {author} {\bibfnamefont {Ali}\
  \bibnamefont {Shakouri}}, \bibinfo {author} {\bibfnamefont {Muhammad~A.}\
  \bibnamefont {Alam}}, \ and\ \bibinfo {author} {\bibfnamefont {Peide~D.}\
  \bibnamefont {Ye}},\ }\bibfield  {title} {\enquote {\bibinfo {title}
  {Steep-slope hysteresis-free negative capacitance mos2 transistors},}\ }\href
  {\doibase 10.1038/s41565-017-0010-1} {\bibfield  {journal} {\bibinfo
  {journal} {Nature Nanotechnology}\ }\textbf {\bibinfo {volume} {13}},\
  \bibinfo {pages} {24--28} (\bibinfo {year} {2018})}\BibitemShut {NoStop}%
\bibitem [{\citenamefont {Lee}\ \emph {et~al.}(2015)\citenamefont {Lee},
  \citenamefont {Chen}, \citenamefont {Liu}, \citenamefont {Chu}, \citenamefont
  {Cheng}, \citenamefont {Xie}, \citenamefont {Liu}, \citenamefont {Lee},
  \citenamefont {Huang}, \citenamefont {Liao}, \citenamefont {Tang},
  \citenamefont {Li},\ and\ \citenamefont {Chen}}]{7409759}%
  \BibitemOpen
  \bibfield  {author} {\bibinfo {author} {\bibfnamefont {M.~H.}\ \bibnamefont
  {Lee}}, \bibinfo {author} {\bibfnamefont {P.-G.}\ \bibnamefont {Chen}},
  \bibinfo {author} {\bibfnamefont {C.}~\bibnamefont {Liu}}, \bibinfo {author}
  {\bibfnamefont {K-Y.}\ \bibnamefont {Chu}}, \bibinfo {author} {\bibfnamefont
  {C.-C.}\ \bibnamefont {Cheng}}, \bibinfo {author} {\bibfnamefont {M.-J.}\
  \bibnamefont {Xie}}, \bibinfo {author} {\bibfnamefont {S.-N.}\ \bibnamefont
  {Liu}}, \bibinfo {author} {\bibfnamefont {J.-W.}\ \bibnamefont {Lee}},
  \bibinfo {author} {\bibfnamefont {S.-J.}\ \bibnamefont {Huang}}, \bibinfo
  {author} {\bibfnamefont {M.-H.}\ \bibnamefont {Liao}}, \bibinfo {author}
  {\bibfnamefont {M.}~\bibnamefont {Tang}}, \bibinfo {author} {\bibfnamefont
  {K.-S.}\ \bibnamefont {Li}}, \ and\ \bibinfo {author} {\bibfnamefont {M.-C.}\
  \bibnamefont {Chen}},\ }\bibfield  {title} {\enquote {\bibinfo {title}
  {Prospects for ferroelectric hfzrox fets with experimentally cet=0.98nm,
  ssfor=42mv/dec, ssrev=28mv/dec, switch-off lt;0.2v, and hysteresis-free
  strategies},}\ }\href {\doibase 10.1109/IEDM.2015.7409759} {\bibfield
  {journal} {\bibinfo  {journal} {2015 IEEE International Electron Devices
  Meeting (IEDM)}\ ,\ \bibinfo {pages} {22.5.1--22.5.4}} (\bibinfo {year}
  {2015})}\BibitemShut {NoStop}%
\bibitem [{\citenamefont {Li}\ \emph {et~al.}(2015)\citenamefont {Li},
  \citenamefont {Chen}, \citenamefont {Lai}, \citenamefont {Lin}, \citenamefont
  {Cheng}, \citenamefont {Chen}, \citenamefont {Wei}, \citenamefont {Hou},
  \citenamefont {Liao}, \citenamefont {Lee}, \citenamefont {Chen},
  \citenamefont {Sheih}, \citenamefont {Yeh}, \citenamefont {Yang},
  \citenamefont {Salahuddin},\ and\ \citenamefont {Hu}}]{7409760}%
  \BibitemOpen
  \bibfield  {author} {\bibinfo {author} {\bibfnamefont {Kai-Shin}\
  \bibnamefont {Li}}, \bibinfo {author} {\bibfnamefont {Pin-Guang}\
  \bibnamefont {Chen}}, \bibinfo {author} {\bibfnamefont {Tung-Yan}\
  \bibnamefont {Lai}}, \bibinfo {author} {\bibfnamefont {Chang-Hsien}\
  \bibnamefont {Lin}}, \bibinfo {author} {\bibfnamefont {Cheng-Chih}\
  \bibnamefont {Cheng}}, \bibinfo {author} {\bibfnamefont {Chun-Chi}\
  \bibnamefont {Chen}}, \bibinfo {author} {\bibfnamefont {Yun-Jie}\
  \bibnamefont {Wei}}, \bibinfo {author} {\bibfnamefont {Yun-Fang}\
  \bibnamefont {Hou}}, \bibinfo {author} {\bibfnamefont {Ming-Han}\
  \bibnamefont {Liao}}, \bibinfo {author} {\bibfnamefont {Min-Hung}\
  \bibnamefont {Lee}}, \bibinfo {author} {\bibfnamefont {Min-Cheng}\
  \bibnamefont {Chen}}, \bibinfo {author} {\bibfnamefont {Jia-Min}\
  \bibnamefont {Sheih}}, \bibinfo {author} {\bibfnamefont {Wen-Kuan}\
  \bibnamefont {Yeh}}, \bibinfo {author} {\bibfnamefont {Fu-Liang}\
  \bibnamefont {Yang}}, \bibinfo {author} {\bibfnamefont {Sayeef}\ \bibnamefont
  {Salahuddin}}, \ and\ \bibinfo {author} {\bibfnamefont {Chenming}\
  \bibnamefont {Hu}},\ }\bibfield  {title} {\enquote {\bibinfo {title}
  {Sub-60mv-swing negative-capacitance finfet without hysteresis},}\ }\href
  {\doibase 10.1109/IEDM.2015.7409760} {\bibfield  {journal} {\bibinfo
  {journal} {2015 IEEE International Electron Devices Meeting (IEDM)}\ ,\
  \bibinfo {pages} {22.6.1--22.6.4}} (\bibinfo {year} {2015})}\BibitemShut
  {NoStop}%
\bibitem [{\citenamefont {Khan}\ \emph {et~al.}(2016)\citenamefont {Khan},
  \citenamefont {Chatterjee}, \citenamefont {Duarte}, \citenamefont {Lu},
  \citenamefont {Sachid}, \citenamefont {Khandelwal}, \citenamefont {Ramesh},
  \citenamefont {Hu},\ and\ \citenamefont {Salahuddin}}]{7331265}%
  \BibitemOpen
  \bibfield  {author} {\bibinfo {author} {\bibfnamefont {Asif~Islam}\
  \bibnamefont {Khan}}, \bibinfo {author} {\bibfnamefont {Korok}\ \bibnamefont
  {Chatterjee}}, \bibinfo {author} {\bibfnamefont {Juan~Pablo}\ \bibnamefont
  {Duarte}}, \bibinfo {author} {\bibfnamefont {Zhongyuan}\ \bibnamefont {Lu}},
  \bibinfo {author} {\bibfnamefont {Angada}\ \bibnamefont {Sachid}}, \bibinfo
  {author} {\bibfnamefont {Sourabh}\ \bibnamefont {Khandelwal}}, \bibinfo
  {author} {\bibfnamefont {Ramamoorthy}\ \bibnamefont {Ramesh}}, \bibinfo
  {author} {\bibfnamefont {Chenming}\ \bibnamefont {Hu}}, \ and\ \bibinfo
  {author} {\bibfnamefont {Sayeef}\ \bibnamefont {Salahuddin}},\ }\bibfield
  {title} {\enquote {\bibinfo {title} {Negative capacitance in short-channel
  finfets externally connected to an epitaxial ferroelectric capacitor},}\
  }\href {\doibase 10.1109/LED.2015.2501319} {\bibfield  {journal} {\bibinfo
  {journal} {IEEE Electron Device Letters}\ }\textbf {\bibinfo {volume} {37}},\
  \bibinfo {pages} {111--114} (\bibinfo {year} {2016})}\BibitemShut {NoStop}%
\bibitem [{\citenamefont {Radisavljevic}\ \emph {et~al.}(2011)\citenamefont
  {Radisavljevic}, \citenamefont {Radenovic}, \citenamefont {Brivio},
  \citenamefont {Giacometti},\ and\ \citenamefont {Kis}}]{Radisavljevic2011}%
  \BibitemOpen
  \bibfield  {author} {\bibinfo {author} {\bibfnamefont {B.}~\bibnamefont
  {Radisavljevic}}, \bibinfo {author} {\bibfnamefont {A.}~\bibnamefont
  {Radenovic}}, \bibinfo {author} {\bibfnamefont {J.}~\bibnamefont {Brivio}},
  \bibinfo {author} {\bibfnamefont {V.}~\bibnamefont {Giacometti}}, \ and\
  \bibinfo {author} {\bibfnamefont {A.}~\bibnamefont {Kis}},\ }\bibfield
  {title} {\enquote {\bibinfo {title} {Single-layer mos2 transistors},}\ }\href
  {\doibase 10.1038/nnano.2010.279} {\bibfield  {journal} {\bibinfo  {journal}
  {Nature Nanotechnology}\ }\textbf {\bibinfo {volume} {6}},\ \bibinfo {pages}
  {147--150} (\bibinfo {year} {2011})}\BibitemShut {NoStop}%
\bibitem [{\citenamefont {Roy}\ \emph {et~al.}(2014)\citenamefont {Roy},
  \citenamefont {Tosun}, \citenamefont {Kang}, \citenamefont {Sachid},
  \citenamefont {Desai}, \citenamefont {Hettick}, \citenamefont {Hu},\ and\
  \citenamefont {Javey}}]{Roy_Nano}%
  \BibitemOpen
  \bibfield  {author} {\bibinfo {author} {\bibfnamefont {Tania}\ \bibnamefont
  {Roy}}, \bibinfo {author} {\bibfnamefont {Mahmut}\ \bibnamefont {Tosun}},
  \bibinfo {author} {\bibfnamefont {Jeong~Seuk}\ \bibnamefont {Kang}}, \bibinfo
  {author} {\bibfnamefont {Angada~B.}\ \bibnamefont {Sachid}}, \bibinfo
  {author} {\bibfnamefont {Sujay~B.}\ \bibnamefont {Desai}}, \bibinfo {author}
  {\bibfnamefont {Mark}\ \bibnamefont {Hettick}}, \bibinfo {author}
  {\bibfnamefont {Chenming~C.}\ \bibnamefont {Hu}}, \ and\ \bibinfo {author}
  {\bibfnamefont {Ali}\ \bibnamefont {Javey}},\ }\bibfield  {title} {\enquote
  {\bibinfo {title} {Field-effect transistors built from all two-dimensional
  material components},}\ }\href {\doibase 10.1021/nn501723y} {\bibfield
  {journal} {\bibinfo  {journal} {ACS Nano}\ }\textbf {\bibinfo {volume} {8}},\
  \bibinfo {pages} {6259--6264} (\bibinfo {year} {2014})}\BibitemShut {NoStop}%
\bibitem [{\citenamefont {Desai}\ \emph {et~al.}(2016)\citenamefont {Desai},
  \citenamefont {Madhvapathy}, \citenamefont {Sachid}, \citenamefont {Llinas},
  \citenamefont {Wang}, \citenamefont {Ahn}, \citenamefont {Pitner},
  \citenamefont {Kim}, \citenamefont {Bokor}, \citenamefont {Hu}, \citenamefont
  {Wong},\ and\ \citenamefont {Javey}}]{Desai99}%
  \BibitemOpen
  \bibfield  {author} {\bibinfo {author} {\bibfnamefont {Sujay~B.}\
  \bibnamefont {Desai}}, \bibinfo {author} {\bibfnamefont {Surabhi~R.}\
  \bibnamefont {Madhvapathy}}, \bibinfo {author} {\bibfnamefont {Angada~B.}\
  \bibnamefont {Sachid}}, \bibinfo {author} {\bibfnamefont {Juan~Pablo}\
  \bibnamefont {Llinas}}, \bibinfo {author} {\bibfnamefont {Qingxiao}\
  \bibnamefont {Wang}}, \bibinfo {author} {\bibfnamefont {Geun~Ho}\
  \bibnamefont {Ahn}}, \bibinfo {author} {\bibfnamefont {Gregory}\ \bibnamefont
  {Pitner}}, \bibinfo {author} {\bibfnamefont {Moon~J.}\ \bibnamefont {Kim}},
  \bibinfo {author} {\bibfnamefont {Jeffrey}\ \bibnamefont {Bokor}}, \bibinfo
  {author} {\bibfnamefont {Chenming}\ \bibnamefont {Hu}}, \bibinfo {author}
  {\bibfnamefont {H.-S.~Philip}\ \bibnamefont {Wong}}, \ and\ \bibinfo {author}
  {\bibfnamefont {Ali}\ \bibnamefont {Javey}},\ }\bibfield  {title} {\enquote
  {\bibinfo {title} {Mos2 transistors with 1-nanometer gate lengths},}\ }\href
  {\doibase 10.1126/science.aah4698} {\bibfield  {journal} {\bibinfo  {journal}
  {Science}\ }\textbf {\bibinfo {volume} {354}},\ \bibinfo {pages} {99--102}
  (\bibinfo {year} {2016})}\BibitemShut {NoStop}%
\bibitem [{\citenamefont {Li}\ \emph {et~al.}(2020)\citenamefont {Li},
  \citenamefont {Tu}, \citenamefont {Sun}, \citenamefont {Fu}, \citenamefont
  {Yu}, \citenamefont {Xing}, \citenamefont {Wang}, \citenamefont {Wang},
  \citenamefont {Jia}, \citenamefont {Wu}, \citenamefont {Tan}, \citenamefont
  {Liang}, \citenamefont {Zhang}, \citenamefont {Zhang}, \citenamefont {Dai},
  \citenamefont {Qiu}, \citenamefont {Li}, \citenamefont {Huang}, \citenamefont
  {Jiao}, \citenamefont {Lai}, \citenamefont {Yan}, \citenamefont {Gao},\ and\
  \citenamefont {Peng}}]{Li2020}%
  \BibitemOpen
  \bibfield  {author} {\bibinfo {author} {\bibfnamefont {Tianran}\ \bibnamefont
  {Li}}, \bibinfo {author} {\bibfnamefont {Teng}\ \bibnamefont {Tu}}, \bibinfo
  {author} {\bibfnamefont {Yuanwei}\ \bibnamefont {Sun}}, \bibinfo {author}
  {\bibfnamefont {Huixia}\ \bibnamefont {Fu}}, \bibinfo {author} {\bibfnamefont
  {Jia}\ \bibnamefont {Yu}}, \bibinfo {author} {\bibfnamefont {Lei}\
  \bibnamefont {Xing}}, \bibinfo {author} {\bibfnamefont {Ziang}\ \bibnamefont
  {Wang}}, \bibinfo {author} {\bibfnamefont {Huimin}\ \bibnamefont {Wang}},
  \bibinfo {author} {\bibfnamefont {Rundong}\ \bibnamefont {Jia}}, \bibinfo
  {author} {\bibfnamefont {Jinxiong}\ \bibnamefont {Wu}}, \bibinfo {author}
  {\bibfnamefont {Congwei}\ \bibnamefont {Tan}}, \bibinfo {author}
  {\bibfnamefont {Yan}\ \bibnamefont {Liang}}, \bibinfo {author} {\bibfnamefont
  {Yichi}\ \bibnamefont {Zhang}}, \bibinfo {author} {\bibfnamefont {Congcong}\
  \bibnamefont {Zhang}}, \bibinfo {author} {\bibfnamefont {Yumin}\ \bibnamefont
  {Dai}}, \bibinfo {author} {\bibfnamefont {Chenguang}\ \bibnamefont {Qiu}},
  \bibinfo {author} {\bibfnamefont {Ming}\ \bibnamefont {Li}}, \bibinfo
  {author} {\bibfnamefont {Ru}~\bibnamefont {Huang}}, \bibinfo {author}
  {\bibfnamefont {Liying}\ \bibnamefont {Jiao}}, \bibinfo {author}
  {\bibfnamefont {Keji}\ \bibnamefont {Lai}}, \bibinfo {author} {\bibfnamefont
  {Binghai}\ \bibnamefont {Yan}}, \bibinfo {author} {\bibfnamefont {Peng}\
  \bibnamefont {Gao}}, \ and\ \bibinfo {author} {\bibfnamefont {Hailin}\
  \bibnamefont {Peng}},\ }\bibfield  {title} {\enquote {\bibinfo {title} {A
  native oxide high-$\kappa$ gate dielectric for two-dimensional
  electronics},}\ }\href {\doibase 10.1038/s41928-020-0444-6} {\bibfield
  {journal} {\bibinfo  {journal} {Nature Electronics}\ }\textbf {\bibinfo
  {volume} {3}},\ \bibinfo {pages} {473--478} (\bibinfo {year}
  {2020})}\BibitemShut {NoStop}%
\bibitem [{\citenamefont {Wu}\ \emph {et~al.}(2022)\citenamefont {Wu},
  \citenamefont {Tian}, \citenamefont {Shen}, \citenamefont {Hou},
  \citenamefont {Ren}, \citenamefont {Gou}, \citenamefont {Sun}, \citenamefont
  {Yang},\ and\ \citenamefont {Ren}}]{Wu2022}%
  \BibitemOpen
  \bibfield  {author} {\bibinfo {author} {\bibfnamefont {Fan}\ \bibnamefont
  {Wu}}, \bibinfo {author} {\bibfnamefont {He}~\bibnamefont {Tian}}, \bibinfo
  {author} {\bibfnamefont {Yang}\ \bibnamefont {Shen}}, \bibinfo {author}
  {\bibfnamefont {Zhan}\ \bibnamefont {Hou}}, \bibinfo {author} {\bibfnamefont
  {Jie}\ \bibnamefont {Ren}}, \bibinfo {author} {\bibfnamefont {Guangyang}\
  \bibnamefont {Gou}}, \bibinfo {author} {\bibfnamefont {Yabin}\ \bibnamefont
  {Sun}}, \bibinfo {author} {\bibfnamefont {Yi}~\bibnamefont {Yang}}, \ and\
  \bibinfo {author} {\bibfnamefont {Tian-Ling}\ \bibnamefont {Ren}},\
  }\bibfield  {title} {\enquote {\bibinfo {title} {Vertical mos2 transistors
  with sub-1-nm gate lengths},}\ }\href {\doibase 10.1038/s41586-021-04323-3}
  {\bibfield  {journal} {\bibinfo  {journal} {Nature}\ }\textbf {\bibinfo
  {volume} {603}},\ \bibinfo {pages} {259--264} (\bibinfo {year}
  {2022})}\BibitemShut {NoStop}%
\bibitem [{\citenamefont {Zou}\ \emph {et~al.}(2020)\citenamefont {Zou},
  \citenamefont {Liu}, \citenamefont {Xu}, \citenamefont {Wang},\ and\
  \citenamefont {Tang}}]{Zou2020}%
  \BibitemOpen
  \bibfield  {author} {\bibinfo {author} {\bibfnamefont {Xiao}\ \bibnamefont
  {Zou}}, \bibinfo {author} {\bibfnamefont {Lu}~\bibnamefont {Liu}}, \bibinfo
  {author} {\bibfnamefont {Jingping}\ \bibnamefont {Xu}}, \bibinfo {author}
  {\bibfnamefont {Hongjiu}\ \bibnamefont {Wang}}, \ and\ \bibinfo {author}
  {\bibfnamefont {Wing-Man}\ \bibnamefont {Tang}},\ }\bibfield  {title}
  {\enquote {\bibinfo {title} {Few-layered mos2 field-effect transistors with a
  vertical channel of sub-10 nm},}\ }\href {\doibase 10.1021/acsami.0c09060}
  {\bibfield  {journal} {\bibinfo  {journal} {ACS Applied Materials \&
  Interfaces}\ }\textbf {\bibinfo {volume} {12}},\ \bibinfo {pages}
  {32943--32950} (\bibinfo {year} {2020})}\BibitemShut {NoStop}%
\bibitem [{\citenamefont {Lemme}\ \emph {et~al.}(2022)\citenamefont {Lemme},
  \citenamefont {Akinwande}, \citenamefont {Huyghebaert},\ and\ \citenamefont
  {Stampfer}}]{Lemme2022}%
  \BibitemOpen
  \bibfield  {author} {\bibinfo {author} {\bibfnamefont {Max~C.}\ \bibnamefont
  {Lemme}}, \bibinfo {author} {\bibfnamefont {Deji}\ \bibnamefont {Akinwande}},
  \bibinfo {author} {\bibfnamefont {Cedric}\ \bibnamefont {Huyghebaert}}, \
  and\ \bibinfo {author} {\bibfnamefont {Christoph}\ \bibnamefont {Stampfer}},\
  }\bibfield  {title} {\enquote {\bibinfo {title} {2d materials for future
  heterogeneous electronics},}\ }\href {\doibase 10.1038/s41467-022-29001-4}
  {\bibfield  {journal} {\bibinfo  {journal} {Nature Communications}\ }\textbf
  {\bibinfo {volume} {13}},\ \bibinfo {pages} {1392} (\bibinfo {year}
  {2022})}\BibitemShut {NoStop}%
\bibitem [{\citenamefont {Novoselov}\ \emph {et~al.}(2004)\citenamefont
  {Novoselov}, \citenamefont {Geim}, \citenamefont {Morozov}, \citenamefont
  {Jiang}, \citenamefont {Zhang}, \citenamefont {Dubonos}, \citenamefont
  {Grigorieva},\ and\ \citenamefont {Firsov}}]{carbon}%
  \BibitemOpen
  \bibfield  {author} {\bibinfo {author} {\bibfnamefont {K.~S.}\ \bibnamefont
  {Novoselov}}, \bibinfo {author} {\bibfnamefont {A.~K.}\ \bibnamefont {Geim}},
  \bibinfo {author} {\bibfnamefont {S.~V.}\ \bibnamefont {Morozov}}, \bibinfo
  {author} {\bibfnamefont {D.}~\bibnamefont {Jiang}}, \bibinfo {author}
  {\bibfnamefont {Y.}~\bibnamefont {Zhang}}, \bibinfo {author} {\bibfnamefont
  {S.~V.}\ \bibnamefont {Dubonos}}, \bibinfo {author} {\bibfnamefont {I.~V.}\
  \bibnamefont {Grigorieva}}, \ and\ \bibinfo {author} {\bibfnamefont {A.~A.}\
  \bibnamefont {Firsov}},\ }\bibfield  {title} {\enquote {\bibinfo {title}
  {Electric field effect in atomically thin carbon films},}\ }\href {\doibase
  10.1126/science.1102896} {\bibfield  {journal} {\bibinfo  {journal} {American
  Association for the Advancement of Science}\ }\textbf {\bibinfo {volume}
  {306}},\ \bibinfo {pages} {666--669} (\bibinfo {year} {2004})}\BibitemShut
  {NoStop}%
\bibitem [{\citenamefont {Bhimanapati}\ \emph {et~al.}(2015)\citenamefont
  {Bhimanapati}, \citenamefont {Lin}, \citenamefont {Meunier}, \citenamefont
  {Jung}, \citenamefont {Cha}, \citenamefont {Das}, \citenamefont {Xiao},
  \citenamefont {Son}, \citenamefont {Strano}, \citenamefont {Cooper},
  \citenamefont {Liang}, \citenamefont {Louie}, \citenamefont {Ringe},
  \citenamefont {Zhou}, \citenamefont {Kim}, \citenamefont {Naik},
  \citenamefont {Sumpter}, \citenamefont {Terrones}, \citenamefont {Xia},
  \citenamefont {Wang}, \citenamefont {Zhu}, \citenamefont {Akinwande},
  \citenamefont {Alem}, \citenamefont {Schuller}, \citenamefont {Schaak},
  \citenamefont {Terrones},\ and\ \citenamefont {Robinson}}]{Ganesh_Nano}%
  \BibitemOpen
  \bibfield  {author} {\bibinfo {author} {\bibfnamefont {Ganesh~R.}\
  \bibnamefont {Bhimanapati}}, \bibinfo {author} {\bibfnamefont {Zhong}\
  \bibnamefont {Lin}}, \bibinfo {author} {\bibfnamefont {Vincent}\ \bibnamefont
  {Meunier}}, \bibinfo {author} {\bibfnamefont {Yeonwoong}\ \bibnamefont
  {Jung}}, \bibinfo {author} {\bibfnamefont {Judy}\ \bibnamefont {Cha}},
  \bibinfo {author} {\bibfnamefont {Saptarshi}\ \bibnamefont {Das}}, \bibinfo
  {author} {\bibfnamefont {Di}~\bibnamefont {Xiao}}, \bibinfo {author}
  {\bibfnamefont {Youngwoo}\ \bibnamefont {Son}}, \bibinfo {author}
  {\bibfnamefont {Michael~S.}\ \bibnamefont {Strano}}, \bibinfo {author}
  {\bibfnamefont {Valentino~R.}\ \bibnamefont {Cooper}}, \bibinfo {author}
  {\bibfnamefont {Liangbo}\ \bibnamefont {Liang}}, \bibinfo {author}
  {\bibfnamefont {Steven~G.}\ \bibnamefont {Louie}}, \bibinfo {author}
  {\bibfnamefont {Emilie}\ \bibnamefont {Ringe}}, \bibinfo {author}
  {\bibfnamefont {Wu}~\bibnamefont {Zhou}}, \bibinfo {author} {\bibfnamefont
  {Steve~S.}\ \bibnamefont {Kim}}, \bibinfo {author} {\bibfnamefont
  {Rajesh~R.}\ \bibnamefont {Naik}}, \bibinfo {author} {\bibfnamefont
  {Bobby~G.}\ \bibnamefont {Sumpter}}, \bibinfo {author} {\bibfnamefont
  {Humberto}\ \bibnamefont {Terrones}}, \bibinfo {author} {\bibfnamefont
  {Fengnian}\ \bibnamefont {Xia}}, \bibinfo {author} {\bibfnamefont {Yeliang}\
  \bibnamefont {Wang}}, \bibinfo {author} {\bibfnamefont {Jun}\ \bibnamefont
  {Zhu}}, \bibinfo {author} {\bibfnamefont {Deji}\ \bibnamefont {Akinwande}},
  \bibinfo {author} {\bibfnamefont {Nasim}\ \bibnamefont {Alem}}, \bibinfo
  {author} {\bibfnamefont {Jon~A.}\ \bibnamefont {Schuller}}, \bibinfo {author}
  {\bibfnamefont {Raymond~E.}\ \bibnamefont {Schaak}}, \bibinfo {author}
  {\bibfnamefont {Mauricio}\ \bibnamefont {Terrones}}, \ and\ \bibinfo {author}
  {\bibfnamefont {Joshua~A.}\ \bibnamefont {Robinson}},\ }\bibfield  {title}
  {\enquote {\bibinfo {title} {Recent advances in two-dimensional materials
  beyond graphene},}\ }\href {\doibase 10.1021/acsnano.5b05556} {\bibfield
  {journal} {\bibinfo  {journal} {ACS Nano}\ }\textbf {\bibinfo {volume} {9}},\
  \bibinfo {pages} {11509--11539} (\bibinfo {year} {2015})}\BibitemShut
  {NoStop}%
\bibitem [{\citenamefont {Butler}\ \emph {et~al.}(2013)\citenamefont {Butler},
  \citenamefont {Hollen}, \citenamefont {Cao}, \citenamefont {Cui},
  \citenamefont {Gupta}, \citenamefont {Gutiérrez}, \citenamefont {Heinz},
  \citenamefont {Hong}, \citenamefont {Huang}, \citenamefont {Ismach},
  \citenamefont {Johnston-Halperin}, \citenamefont {Kuno}, \citenamefont
  {Plashnitsa}, \citenamefont {Robinson}, \citenamefont {Ruoff}, \citenamefont
  {Salahuddin}, \citenamefont {Shan}, \citenamefont {Shi}, \citenamefont
  {Spencer}, \citenamefont {Terrones}, \citenamefont {Windl},\ and\
  \citenamefont {Goldberger}}]{Butler_Nano}%
  \BibitemOpen
  \bibfield  {author} {\bibinfo {author} {\bibfnamefont {Sheneve~Z.}\
  \bibnamefont {Butler}}, \bibinfo {author} {\bibfnamefont {Shawna~M.}\
  \bibnamefont {Hollen}}, \bibinfo {author} {\bibfnamefont {Linyou}\
  \bibnamefont {Cao}}, \bibinfo {author} {\bibfnamefont {Yi}~\bibnamefont
  {Cui}}, \bibinfo {author} {\bibfnamefont {Jay~A.}\ \bibnamefont {Gupta}},
  \bibinfo {author} {\bibfnamefont {Humberto~R.}\ \bibnamefont {Gutiérrez}},
  \bibinfo {author} {\bibfnamefont {Tony~F.}\ \bibnamefont {Heinz}}, \bibinfo
  {author} {\bibfnamefont {Seung~Sae}\ \bibnamefont {Hong}}, \bibinfo {author}
  {\bibfnamefont {Jiaxing}\ \bibnamefont {Huang}}, \bibinfo {author}
  {\bibfnamefont {Ariel~F.}\ \bibnamefont {Ismach}}, \bibinfo {author}
  {\bibfnamefont {Ezekiel}\ \bibnamefont {Johnston-Halperin}}, \bibinfo
  {author} {\bibfnamefont {Masaru}\ \bibnamefont {Kuno}}, \bibinfo {author}
  {\bibfnamefont {Vladimir~V.}\ \bibnamefont {Plashnitsa}}, \bibinfo {author}
  {\bibfnamefont {Richard~D.}\ \bibnamefont {Robinson}}, \bibinfo {author}
  {\bibfnamefont {Rodney~S.}\ \bibnamefont {Ruoff}}, \bibinfo {author}
  {\bibfnamefont {Sayeef}\ \bibnamefont {Salahuddin}}, \bibinfo {author}
  {\bibfnamefont {Jie}\ \bibnamefont {Shan}}, \bibinfo {author} {\bibfnamefont
  {Li}~\bibnamefont {Shi}}, \bibinfo {author} {\bibfnamefont {Michael~G.}\
  \bibnamefont {Spencer}}, \bibinfo {author} {\bibfnamefont {Mauricio}\
  \bibnamefont {Terrones}}, \bibinfo {author} {\bibfnamefont {Wolfgang}\
  \bibnamefont {Windl}}, \ and\ \bibinfo {author} {\bibfnamefont {Joshua~E.}\
  \bibnamefont {Goldberger}},\ }\bibfield  {title} {\enquote {\bibinfo {title}
  {Progress, challenges, and opportunities in two-dimensional materials beyond
  graphene},}\ }\href {\doibase 10.1021/nn400280c} {\bibfield  {journal}
  {\bibinfo  {journal} {ACS Nano}\ }\textbf {\bibinfo {volume} {7}},\ \bibinfo
  {pages} {2898--2926} (\bibinfo {year} {2013})}\BibitemShut {NoStop}%
\bibitem [{\citenamefont {Lin}\ \emph {et~al.}(2018)\citenamefont {Lin},
  \citenamefont {Jariwala}, \citenamefont {Bersch}, \citenamefont {Xu},
  \citenamefont {Nie}, \citenamefont {Wang}, \citenamefont {Eichfeld},
  \citenamefont {Zhang}, \citenamefont {Choudhury}, \citenamefont {Pan},
  \citenamefont {Addou}, \citenamefont {Smyth}, \citenamefont {Li},
  \citenamefont {Zhang}, \citenamefont {Haque}, \citenamefont {Fölsch},
  \citenamefont {Feenstra}, \citenamefont {Wallace}, \citenamefont {Cho},
  \citenamefont {Fullerton-Shirey}, \citenamefont {Redwing},\ and\
  \citenamefont {Robinson}}]{Lin_Nano}%
  \BibitemOpen
  \bibfield  {author} {\bibinfo {author} {\bibfnamefont {Yu-Chuan}\
  \bibnamefont {Lin}}, \bibinfo {author} {\bibfnamefont {Bhakti}\ \bibnamefont
  {Jariwala}}, \bibinfo {author} {\bibfnamefont {Brian~M.}\ \bibnamefont
  {Bersch}}, \bibinfo {author} {\bibfnamefont {Ke}~\bibnamefont {Xu}}, \bibinfo
  {author} {\bibfnamefont {Yifan}\ \bibnamefont {Nie}}, \bibinfo {author}
  {\bibfnamefont {Baoming}\ \bibnamefont {Wang}}, \bibinfo {author}
  {\bibfnamefont {Sarah~M.}\ \bibnamefont {Eichfeld}}, \bibinfo {author}
  {\bibfnamefont {Xiaotian}\ \bibnamefont {Zhang}}, \bibinfo {author}
  {\bibfnamefont {Tanushree~H.}\ \bibnamefont {Choudhury}}, \bibinfo {author}
  {\bibfnamefont {Yi}~\bibnamefont {Pan}}, \bibinfo {author} {\bibfnamefont
  {Rafik}\ \bibnamefont {Addou}}, \bibinfo {author} {\bibfnamefont
  {Christopher~M.}\ \bibnamefont {Smyth}}, \bibinfo {author} {\bibfnamefont
  {Jun}\ \bibnamefont {Li}}, \bibinfo {author} {\bibfnamefont {Kehao}\
  \bibnamefont {Zhang}}, \bibinfo {author} {\bibfnamefont {M.~Aman}\
  \bibnamefont {Haque}}, \bibinfo {author} {\bibfnamefont {Stefan}\
  \bibnamefont {Fölsch}}, \bibinfo {author} {\bibfnamefont {Randall~M.}\
  \bibnamefont {Feenstra}}, \bibinfo {author} {\bibfnamefont {Robert~M.}\
  \bibnamefont {Wallace}}, \bibinfo {author} {\bibfnamefont {Kyeongjae}\
  \bibnamefont {Cho}}, \bibinfo {author} {\bibfnamefont {Susan~K.}\
  \bibnamefont {Fullerton-Shirey}}, \bibinfo {author} {\bibfnamefont {Joan~M.}\
  \bibnamefont {Redwing}}, \ and\ \bibinfo {author} {\bibfnamefont {Joshua~A.}\
  \bibnamefont {Robinson}},\ }\bibfield  {title} {\enquote {\bibinfo {title}
  {Realizing large-scale, electronic-grade two-dimensional semiconductors},}\
  }\href {\doibase 10.1021/acsnano.7b07059} {\bibfield  {journal} {\bibinfo
  {journal} {ACS Nano}\ }\textbf {\bibinfo {volume} {12}},\ \bibinfo {pages}
  {965--975} (\bibinfo {year} {2018})}\BibitemShut {NoStop}%
\bibitem [{\citenamefont {Cai}\ \emph {et~al.}(2018)\citenamefont {Cai},
  \citenamefont {Liu}, \citenamefont {Zou},\ and\ \citenamefont
  {Cheng}}]{Cai_CR}%
  \BibitemOpen
  \bibfield  {author} {\bibinfo {author} {\bibfnamefont {Zhengyang}\
  \bibnamefont {Cai}}, \bibinfo {author} {\bibfnamefont {Bilu}\ \bibnamefont
  {Liu}}, \bibinfo {author} {\bibfnamefont {Xiaolong}\ \bibnamefont {Zou}}, \
  and\ \bibinfo {author} {\bibfnamefont {Hui-Ming}\ \bibnamefont {Cheng}},\
  }\bibfield  {title} {\enquote {\bibinfo {title} {Chemical vapor deposition
  growth and applications of two-dimensional materials and their
  heterostructures},}\ }\href {\doibase 10.1021/acs.chemrev.7b00536} {\bibfield
   {journal} {\bibinfo  {journal} {Chemical Reviews}\ }\textbf {\bibinfo
  {volume} {118}},\ \bibinfo {pages} {6091--6133} (\bibinfo {year}
  {2018})}\BibitemShut {NoStop}%
\bibitem [{\citenamefont {Dumcenco}\ \emph {et~al.}(2015)\citenamefont
  {Dumcenco}, \citenamefont {Ovchinnikov}, \citenamefont {Marinov},
  \citenamefont {Lazić}, \citenamefont {Gibertini}, \citenamefont {Marzari},
  \citenamefont {Sanchez}, \citenamefont {Kung}, \citenamefont {Krasnozhon},
  \citenamefont {Chen}, \citenamefont {Bertolazzi}, \citenamefont {Gillet},
  \citenamefont {Fontcuberta~i Morral}, \citenamefont {Radenovic},\ and\
  \citenamefont {Kis}}]{DD_Nano}%
  \BibitemOpen
  \bibfield  {author} {\bibinfo {author} {\bibfnamefont {Dumitru}\ \bibnamefont
  {Dumcenco}}, \bibinfo {author} {\bibfnamefont {Dmitry}\ \bibnamefont
  {Ovchinnikov}}, \bibinfo {author} {\bibfnamefont {Kolyo}\ \bibnamefont
  {Marinov}}, \bibinfo {author} {\bibfnamefont {Predrag}\ \bibnamefont
  {Lazić}}, \bibinfo {author} {\bibfnamefont {Marco}\ \bibnamefont
  {Gibertini}}, \bibinfo {author} {\bibfnamefont {Nicola}\ \bibnamefont
  {Marzari}}, \bibinfo {author} {\bibfnamefont {Oriol~Lopez}\ \bibnamefont
  {Sanchez}}, \bibinfo {author} {\bibfnamefont {Yen-Cheng}\ \bibnamefont
  {Kung}}, \bibinfo {author} {\bibfnamefont {Daria}\ \bibnamefont
  {Krasnozhon}}, \bibinfo {author} {\bibfnamefont {Ming-Wei}\ \bibnamefont
  {Chen}}, \bibinfo {author} {\bibfnamefont {Simone}\ \bibnamefont
  {Bertolazzi}}, \bibinfo {author} {\bibfnamefont {Philippe}\ \bibnamefont
  {Gillet}}, \bibinfo {author} {\bibfnamefont {Anna}\ \bibnamefont
  {Fontcuberta~i Morral}}, \bibinfo {author} {\bibfnamefont {Aleksandra}\
  \bibnamefont {Radenovic}}, \ and\ \bibinfo {author} {\bibfnamefont {Andras}\
  \bibnamefont {Kis}},\ }\bibfield  {title} {\enquote {\bibinfo {title}
  {Large-area epitaxial monolayer mos2},}\ }\href {\doibase
  10.1021/acsnano.5b01281} {\bibfield  {journal} {\bibinfo  {journal} {ACS
  Nano}\ }\textbf {\bibinfo {volume} {9}},\ \bibinfo {pages} {4611--4620}
  (\bibinfo {year} {2015})}\BibitemShut {NoStop}%
\bibitem [{\citenamefont {Priydarshi}\ \emph {et~al.}(2022)\citenamefont
  {Priydarshi}, \citenamefont {Chauhan}, \citenamefont {Bhowmick},\ and\
  \citenamefont {Agarwal}}]{D2NR02382B}%
  \BibitemOpen
  \bibfield  {author} {\bibinfo {author} {\bibfnamefont {Achintya}\
  \bibnamefont {Priydarshi}}, \bibinfo {author} {\bibfnamefont {Yogesh~Singh}\
  \bibnamefont {Chauhan}}, \bibinfo {author} {\bibfnamefont {Somnath}\
  \bibnamefont {Bhowmick}}, \ and\ \bibinfo {author} {\bibfnamefont {Amit}\
  \bibnamefont {Agarwal}},\ }\bibfield  {title} {\enquote {\bibinfo {title}
  {Large and anisotropic carrier mobility in monolayers of the ma2z4 series (m
  = cr{,} mo{,} w; a = si{,} ge; and z = n{,} p)},}\ }\href {\doibase
  10.1039/D2NR02382B} {\bibfield  {journal} {\bibinfo  {journal} {Nanoscale}\
  }\textbf {\bibinfo {volume} {14}},\ \bibinfo {pages} {11988--11997} (\bibinfo
  {year} {2022})}\BibitemShut {NoStop}%
\bibitem [{\citenamefont {Sarkar}\ \emph {et~al.}(2020)\citenamefont {Sarkar},
  \citenamefont {Ghosh}, \citenamefont {Singh}, \citenamefont {Bhowmick},
  \citenamefont {Lin}, \citenamefont {Bansil},\ and\ \citenamefont
  {Agarwal}}]{PhysRevB.102.035420}%
  \BibitemOpen
  \bibfield  {author} {\bibinfo {author} {\bibfnamefont {Anan~Bari}\
  \bibnamefont {Sarkar}}, \bibinfo {author} {\bibfnamefont {Barun}\
  \bibnamefont {Ghosh}}, \bibinfo {author} {\bibfnamefont {Bahadur}\
  \bibnamefont {Singh}}, \bibinfo {author} {\bibfnamefont {Somnath}\
  \bibnamefont {Bhowmick}}, \bibinfo {author} {\bibfnamefont {Hsin}\
  \bibnamefont {Lin}}, \bibinfo {author} {\bibfnamefont {Arun}\ \bibnamefont
  {Bansil}}, \ and\ \bibinfo {author} {\bibfnamefont {Amit}\ \bibnamefont
  {Agarwal}},\ }\bibfield  {title} {\enquote {\bibinfo {title}
  {${\mathrm{k}}_{2}{\mathrm{cos}}_{2}$: A two-dimensional in-plane
  antiferromagnetic insulator},}\ }\href {\doibase 10.1103/PhysRevB.102.035420}
  {\bibfield  {journal} {\bibinfo  {journal} {Phys. Rev. B}\ }\textbf {\bibinfo
  {volume} {102}},\ \bibinfo {pages} {035420} (\bibinfo {year}
  {2020})}\BibitemShut {NoStop}%
\bibitem [{\citenamefont {Ghosh}\ \emph {et~al.}(2019)\citenamefont {Ghosh},
  \citenamefont {Mardanya}, \citenamefont {Singh}, \citenamefont {Zhou},
  \citenamefont {Wang}, \citenamefont {Chang}, \citenamefont {Su},
  \citenamefont {Lin}, \citenamefont {Agarwal},\ and\ \citenamefont
  {Bansil}}]{PhysRevB.100.235101}%
  \BibitemOpen
  \bibfield  {author} {\bibinfo {author} {\bibfnamefont {Barun}\ \bibnamefont
  {Ghosh}}, \bibinfo {author} {\bibfnamefont {Sougata}\ \bibnamefont
  {Mardanya}}, \bibinfo {author} {\bibfnamefont {Bahadur}\ \bibnamefont
  {Singh}}, \bibinfo {author} {\bibfnamefont {Xiaoting}\ \bibnamefont {Zhou}},
  \bibinfo {author} {\bibfnamefont {Baokai}\ \bibnamefont {Wang}}, \bibinfo
  {author} {\bibfnamefont {Tay-Rong}\ \bibnamefont {Chang}}, \bibinfo {author}
  {\bibfnamefont {Chenliang}\ \bibnamefont {Su}}, \bibinfo {author}
  {\bibfnamefont {Hsin}\ \bibnamefont {Lin}}, \bibinfo {author} {\bibfnamefont
  {Amit}\ \bibnamefont {Agarwal}}, \ and\ \bibinfo {author} {\bibfnamefont
  {Arun}\ \bibnamefont {Bansil}},\ }\bibfield  {title} {\enquote {\bibinfo
  {title} {Saddle-point van hove singularity and dual topological state in
  ${\mathrm{pt}}_{2}{\mathrm{hgse}}_{3}$},}\ }\href {\doibase
  10.1103/PhysRevB.100.235101} {\bibfield  {journal} {\bibinfo  {journal}
  {Phys. Rev. B}\ }\textbf {\bibinfo {volume} {100}},\ \bibinfo {pages}
  {235101} (\bibinfo {year} {2019})}\BibitemShut {NoStop}%
\bibitem [{\citenamefont {Ghosh}\ \emph {et~al.}(2018)\citenamefont {Ghosh},
  \citenamefont {Puri}, \citenamefont {Agarwal},\ and\ \citenamefont
  {Bhowmick}}]{Ghosh_jpcc}%
  \BibitemOpen
  \bibfield  {author} {\bibinfo {author} {\bibfnamefont {Barun}\ \bibnamefont
  {Ghosh}}, \bibinfo {author} {\bibfnamefont {Shivam}\ \bibnamefont {Puri}},
  \bibinfo {author} {\bibfnamefont {Amit}\ \bibnamefont {Agarwal}}, \ and\
  \bibinfo {author} {\bibfnamefont {Somnath}\ \bibnamefont {Bhowmick}},\
  }\bibfield  {title} {\enquote {\bibinfo {title} {Snp3: A previously
  unexplored two-dimensional material},}\ }\href {\doibase
  10.1021/acs.jpcc.8b06668} {\bibfield  {journal} {\bibinfo  {journal} {The
  Journal of Physical Chemistry C}\ }\textbf {\bibinfo {volume} {122}},\
  \bibinfo {pages} {18185--18191} (\bibinfo {year} {2018})}\BibitemShut
  {NoStop}%
\bibitem [{\citenamefont {Ghosh}\ \emph {et~al.}(2016)\citenamefont {Ghosh},
  \citenamefont {Singh}, \citenamefont {Prasad},\ and\ \citenamefont
  {Agarwal}}]{PhysRevB.94.205426}%
  \BibitemOpen
  \bibfield  {author} {\bibinfo {author} {\bibfnamefont {Barun}\ \bibnamefont
  {Ghosh}}, \bibinfo {author} {\bibfnamefont {Bahadur}\ \bibnamefont {Singh}},
  \bibinfo {author} {\bibfnamefont {R.}~\bibnamefont {Prasad}}, \ and\ \bibinfo
  {author} {\bibfnamefont {Amit}\ \bibnamefont {Agarwal}},\ }\bibfield  {title}
  {\enquote {\bibinfo {title} {Electric-field tunable dirac semimetal state in
  phosphorene thin films},}\ }\href {\doibase 10.1103/PhysRevB.94.205426}
  {\bibfield  {journal} {\bibinfo  {journal} {Phys. Rev. B}\ }\textbf {\bibinfo
  {volume} {94}},\ \bibinfo {pages} {205426} (\bibinfo {year}
  {2016})}\BibitemShut {NoStop}%
\bibitem [{\citenamefont {Oliva}\ \emph {et~al.}(2020)\citenamefont {Oliva},
  \citenamefont {Backman}, \citenamefont {Capua}, \citenamefont {Cavalieri},
  \citenamefont {Luisier},\ and\ \citenamefont {Ionescu}}]{DT-FET}%
  \BibitemOpen
  \bibfield  {author} {\bibinfo {author} {\bibfnamefont {Nicol{\`o}}\
  \bibnamefont {Oliva}}, \bibinfo {author} {\bibfnamefont {Jonathan}\
  \bibnamefont {Backman}}, \bibinfo {author} {\bibfnamefont {Luca}\
  \bibnamefont {Capua}}, \bibinfo {author} {\bibfnamefont {Matteo}\
  \bibnamefont {Cavalieri}}, \bibinfo {author} {\bibfnamefont {Mathieu}\
  \bibnamefont {Luisier}}, \ and\ \bibinfo {author} {\bibfnamefont {Adrian~M.}\
  \bibnamefont {Ionescu}},\ }\bibfield  {title} {\enquote {\bibinfo {title}
  {Wse2/snse2 vdw heterojunction tunnel fet with subthermionic characteristic
  and mosfet co-integrated on same wse2 flake},}\ }\href {\doibase
  10.1038/s41699-020-0142-2} {\bibfield  {journal} {\bibinfo  {journal} {npj 2D
  Materials and Applications}\ }\textbf {\bibinfo {volume} {4}},\ \bibinfo
  {pages} {5} (\bibinfo {year} {2020})}\BibitemShut {NoStop}%
\bibitem [{\citenamefont {Liu}\ \emph {et~al.}(2020{\natexlab{a}})\citenamefont
  {Liu}, \citenamefont {Chen}, \citenamefont {Wang}, \citenamefont {Liu},
  \citenamefont {Jiang}, \citenamefont {Zhang}, \citenamefont {Liu},\ and\
  \citenamefont {Zhou}}]{Liu2020}%
  \BibitemOpen
  \bibfield  {author} {\bibinfo {author} {\bibfnamefont {Chunsen}\ \bibnamefont
  {Liu}}, \bibinfo {author} {\bibfnamefont {Huawei}\ \bibnamefont {Chen}},
  \bibinfo {author} {\bibfnamefont {Shuiyuan}\ \bibnamefont {Wang}}, \bibinfo
  {author} {\bibfnamefont {Qi}~\bibnamefont {Liu}}, \bibinfo {author}
  {\bibfnamefont {Yu-Gang}\ \bibnamefont {Jiang}}, \bibinfo {author}
  {\bibfnamefont {David~Wei}\ \bibnamefont {Zhang}}, \bibinfo {author}
  {\bibfnamefont {Ming}\ \bibnamefont {Liu}}, \ and\ \bibinfo {author}
  {\bibfnamefont {Peng}\ \bibnamefont {Zhou}},\ }\bibfield  {title} {\enquote
  {\bibinfo {title} {Two-dimensional materials for next-generation computing
  technologies},}\ }\href {\doibase 10.1038/s41565-020-0724-3} {\bibfield
  {journal} {\bibinfo  {journal} {Nature Nanotechnology}\ }\textbf {\bibinfo
  {volume} {15}},\ \bibinfo {pages} {545--557} (\bibinfo {year}
  {2020}{\natexlab{a}})}\BibitemShut {NoStop}%
\bibitem [{\citenamefont {Logoteta}\ \emph {et~al.}(2019)\citenamefont
  {Logoteta}, \citenamefont {Pala}, \citenamefont {Choukroun}, \citenamefont
  {Dollfus},\ and\ \citenamefont {Iannaccone}}]{Pala_1}%
  \BibitemOpen
  \bibfield  {author} {\bibinfo {author} {\bibfnamefont {Demetrio}\
  \bibnamefont {Logoteta}}, \bibinfo {author} {\bibfnamefont {Marco~G.}\
  \bibnamefont {Pala}}, \bibinfo {author} {\bibfnamefont {Jean}\ \bibnamefont
  {Choukroun}}, \bibinfo {author} {\bibfnamefont {Philippe}\ \bibnamefont
  {Dollfus}}, \ and\ \bibinfo {author} {\bibfnamefont {Giuseppe}\ \bibnamefont
  {Iannaccone}},\ }\bibfield  {title} {\enquote {\bibinfo {title} {A
  steep-slope {M}o{S}$_2$-nanoribbon mosfet based on an intrinsic cold-contact
  effect},}\ }\href {\doibase 10.1109/LED.2019.2928131} {\bibfield  {journal}
  {\bibinfo  {journal} {IEEE Electron Device Letters}\ }\textbf {\bibinfo
  {volume} {40}},\ \bibinfo {pages} {1550--1553} (\bibinfo {year}
  {2019})}\BibitemShut {NoStop}%
\bibitem [{\citenamefont {Nadeem}\ \emph {et~al.}(2021)\citenamefont {Nadeem},
  \citenamefont {Di~Bernardo}, \citenamefont {Wang}, \citenamefont {Fuhrer},\
  and\ \citenamefont {Culcer}}]{Nadeem}%
  \BibitemOpen
  \bibfield  {author} {\bibinfo {author} {\bibfnamefont {Muhammad}\
  \bibnamefont {Nadeem}}, \bibinfo {author} {\bibfnamefont {Iolanda}\
  \bibnamefont {Di~Bernardo}}, \bibinfo {author} {\bibfnamefont {Xiaolin}\
  \bibnamefont {Wang}}, \bibinfo {author} {\bibfnamefont {Michael~S.}\
  \bibnamefont {Fuhrer}}, \ and\ \bibinfo {author} {\bibfnamefont {Dimitrie}\
  \bibnamefont {Culcer}},\ }\bibfield  {title} {\enquote {\bibinfo {title}
  {Overcoming boltzmann’s tyranny in a transistor via the topological quantum
  field effect},}\ }\href {\doibase 10.1021/acs.nanolett.1c00378} {\bibfield
  {journal} {\bibinfo  {journal} {Nano Letters}\ }\textbf {\bibinfo {volume}
  {21}},\ \bibinfo {pages} {3155--3161} (\bibinfo {year} {2021})}\BibitemShut
  {NoStop}%
\bibitem [{\citenamefont {Qiu}\ \emph {et~al.}(2018)\citenamefont {Qiu},
  \citenamefont {Liu}, \citenamefont {Xu}, \citenamefont {Deng}, \citenamefont
  {Xiao}, \citenamefont {Si}, \citenamefont {Lin}, \citenamefont {Zhang},
  \citenamefont {Wang}, \citenamefont {Guo} \emph {et~al.}}]{qiu2018dirac}%
  \BibitemOpen
  \bibfield  {author} {\bibinfo {author} {\bibfnamefont {Chenguang}\
  \bibnamefont {Qiu}}, \bibinfo {author} {\bibfnamefont {Fei}\ \bibnamefont
  {Liu}}, \bibinfo {author} {\bibfnamefont {Lin}\ \bibnamefont {Xu}}, \bibinfo
  {author} {\bibfnamefont {Bing}\ \bibnamefont {Deng}}, \bibinfo {author}
  {\bibfnamefont {Mengmeng}\ \bibnamefont {Xiao}}, \bibinfo {author}
  {\bibfnamefont {Jia}\ \bibnamefont {Si}}, \bibinfo {author} {\bibfnamefont
  {Li}~\bibnamefont {Lin}}, \bibinfo {author} {\bibfnamefont {Zhiyong}\
  \bibnamefont {Zhang}}, \bibinfo {author} {\bibfnamefont {Jian}\ \bibnamefont
  {Wang}}, \bibinfo {author} {\bibfnamefont {Hong}\ \bibnamefont {Guo}},  \emph
  {et~al.},\ }\bibfield  {title} {\enquote {\bibinfo {title} {Dirac-source
  field-effect transistors as energy-efficient, high-performance electronic
  switches},}\ }\href@noop {} {\bibfield  {journal} {\bibinfo  {journal}
  {Science}\ }\textbf {\bibinfo {volume} {361}},\ \bibinfo {pages} {387--392}
  (\bibinfo {year} {2018})}\BibitemShut {NoStop}%
\bibitem [{\citenamefont {Liu}\ \emph {et~al.}(2020{\natexlab{b}})\citenamefont
  {Liu}, \citenamefont {Jaiswal}, \citenamefont {Shahi}, \citenamefont {Wei},
  \citenamefont {Fu}, \citenamefont {Chang}, \citenamefont {Chakravarty},
  \citenamefont {Yao},\ and\ \citenamefont {Li}}]{Gr_MoS2_DSFET}%
  \BibitemOpen
  \bibfield  {author} {\bibinfo {author} {\bibfnamefont {Maomao}\ \bibnamefont
  {Liu}}, \bibinfo {author} {\bibfnamefont {Hemendra~Nath}\ \bibnamefont
  {Jaiswal}}, \bibinfo {author} {\bibfnamefont {Simran}\ \bibnamefont {Shahi}},
  \bibinfo {author} {\bibfnamefont {Sichen}\ \bibnamefont {Wei}}, \bibinfo
  {author} {\bibfnamefont {Yu}~\bibnamefont {Fu}}, \bibinfo {author}
  {\bibfnamefont {Chaoran}\ \bibnamefont {Chang}}, \bibinfo {author}
  {\bibfnamefont {Anindita}\ \bibnamefont {Chakravarty}}, \bibinfo {author}
  {\bibfnamefont {Fei}\ \bibnamefont {Yao}}, \ and\ \bibinfo {author}
  {\bibfnamefont {Huamin}\ \bibnamefont {Li}},\ }\bibfield  {title} {\enquote
  {\bibinfo {title} {Monolayer {M}o{S}$_2$ steep-slope transistors with
  record-high sub-60-mv/decade current density using dirac-source electron
  injection},}\ }\href {\doibase 10.1109/IEDM13553.2020.9371961} {\bibfield
  {journal} {\bibinfo  {journal} {IEEE International Electron Devices Meeting
  (IEDM)}\ ,\ \bibinfo {pages} {12.5.1--12.5.4}} (\bibinfo {year}
  {2020}{\natexlab{b}})}\BibitemShut {NoStop}%
\bibitem [{\citenamefont {Hong}\ \emph {et~al.}(2020)\citenamefont {Hong},
  \citenamefont {Liu}, \citenamefont {Wang}, \citenamefont {Zhou},
  \citenamefont {Ma}, \citenamefont {Xu}, \citenamefont {Feng}, \citenamefont
  {Chen}, \citenamefont {Chen}, \citenamefont {Sun}, \citenamefont {Chen},
  \citenamefont {Cheng},\ and\ \citenamefont {Ren}}]{Hong}%
  \BibitemOpen
  \bibfield  {author} {\bibinfo {author} {\bibfnamefont {Yi-Lun}\ \bibnamefont
  {Hong}}, \bibinfo {author} {\bibfnamefont {Zhibo}\ \bibnamefont {Liu}},
  \bibinfo {author} {\bibfnamefont {Lei}\ \bibnamefont {Wang}}, \bibinfo
  {author} {\bibfnamefont {Tianya}\ \bibnamefont {Zhou}}, \bibinfo {author}
  {\bibfnamefont {Wei}\ \bibnamefont {Ma}}, \bibinfo {author} {\bibfnamefont
  {Chuan}\ \bibnamefont {Xu}}, \bibinfo {author} {\bibfnamefont {Shun}\
  \bibnamefont {Feng}}, \bibinfo {author} {\bibfnamefont {Long}\ \bibnamefont
  {Chen}}, \bibinfo {author} {\bibfnamefont {Mao-Lin}\ \bibnamefont {Chen}},
  \bibinfo {author} {\bibfnamefont {Dong-Ming}\ \bibnamefont {Sun}}, \bibinfo
  {author} {\bibfnamefont {Xing-Qiu}\ \bibnamefont {Chen}}, \bibinfo {author}
  {\bibfnamefont {Hui-Ming}\ \bibnamefont {Cheng}}, \ and\ \bibinfo {author}
  {\bibfnamefont {Wencai}\ \bibnamefont {Ren}},\ }\bibfield  {title} {\enquote
  {\bibinfo {title} {Chemical vapor deposition of layered two-dimensional
  mosi2n4 materials},}\ }\href {\doibase 10.1126/science.abb7023} {\bibfield
  {journal} {\bibinfo  {journal} {Science}\ }\textbf {\bibinfo {volume}
  {369}},\ \bibinfo {pages} {670--674} (\bibinfo {year} {2020})}\BibitemShut
  {NoStop}%
\bibitem [{\citenamefont {Cao}\ \emph {et~al.}(2021)\citenamefont {Cao},
  \citenamefont {Zhou}, \citenamefont {Wang}, \citenamefont {Ang},\ and\
  \citenamefont {Ang}}]{Cao}%
  \BibitemOpen
  \bibfield  {author} {\bibinfo {author} {\bibfnamefont {Liemao}\ \bibnamefont
  {Cao}}, \bibinfo {author} {\bibfnamefont {Guanghui}\ \bibnamefont {Zhou}},
  \bibinfo {author} {\bibfnamefont {Qianqian}\ \bibnamefont {Wang}}, \bibinfo
  {author} {\bibfnamefont {L.~K.}\ \bibnamefont {Ang}}, \ and\ \bibinfo
  {author} {\bibfnamefont {Yee~Sin}\ \bibnamefont {Ang}},\ }\bibfield  {title}
  {\enquote {\bibinfo {title} {Two-dimensional van der waals electrical contact
  to monolayer mosi2n4},}\ }\href {\doibase 10.1063/5.0033241} {\bibfield
  {journal} {\bibinfo  {journal} {Applied Physics Letters}\ }\textbf {\bibinfo
  {volume} {118}},\ \bibinfo {pages} {013106} (\bibinfo {year}
  {2021})}\BibitemShut {NoStop}%
\bibitem [{\citenamefont {Bafekry}\ \emph {et~al.}(2021)\citenamefont
  {Bafekry}, \citenamefont {Faraji}, \citenamefont {Hoat}, \citenamefont
  {Shahrokhi}, \citenamefont {Fadlallah}, \citenamefont {Shojaei},
  \citenamefont {Feghhi}, \citenamefont {Ghergherehchi},\ and\ \citenamefont
  {Gogova}}]{Bafekry_2021}%
  \BibitemOpen
  \bibfield  {author} {\bibinfo {author} {\bibfnamefont {A}~\bibnamefont
  {Bafekry}}, \bibinfo {author} {\bibfnamefont {M}~\bibnamefont {Faraji}},
  \bibinfo {author} {\bibfnamefont {D~M}\ \bibnamefont {Hoat}}, \bibinfo
  {author} {\bibfnamefont {M}~\bibnamefont {Shahrokhi}}, \bibinfo {author}
  {\bibfnamefont {M~M}\ \bibnamefont {Fadlallah}}, \bibinfo {author}
  {\bibfnamefont {F}~\bibnamefont {Shojaei}}, \bibinfo {author} {\bibfnamefont
  {S~A~H}\ \bibnamefont {Feghhi}}, \bibinfo {author} {\bibfnamefont
  {M}~\bibnamefont {Ghergherehchi}}, \ and\ \bibinfo {author} {\bibfnamefont
  {D}~\bibnamefont {Gogova}},\ }\bibfield  {title} {\enquote {\bibinfo {title}
  {{MoSi}2n4 single-layer: a novel two-dimensional material with outstanding
  mechanical, thermal, electronic and optical properties},}\ }\href {\doibase
  10.1088/1361-6463/abdb6b} {\bibfield  {journal} {\bibinfo  {journal} {Journal
  of Physics D: Applied Physics}\ }\textbf {\bibinfo {volume} {54}},\ \bibinfo
  {pages} {155303} (\bibinfo {year} {2021})}\BibitemShut {NoStop}%
\bibitem [{\citenamefont {Mortazavi}\ \emph {et~al.}(2021)\citenamefont
  {Mortazavi}, \citenamefont {Javvaji}, \citenamefont {Shojaei}, \citenamefont
  {Rabczuk}, \citenamefont {Shapeev},\ and\ \citenamefont
  {Zhuang}}]{MORTAZAVI2021105716}%
  \BibitemOpen
  \bibfield  {author} {\bibinfo {author} {\bibfnamefont {Bohayra}\ \bibnamefont
  {Mortazavi}}, \bibinfo {author} {\bibfnamefont {Brahmanandam}\ \bibnamefont
  {Javvaji}}, \bibinfo {author} {\bibfnamefont {Fazel}\ \bibnamefont
  {Shojaei}}, \bibinfo {author} {\bibfnamefont {Timon}\ \bibnamefont
  {Rabczuk}}, \bibinfo {author} {\bibfnamefont {Alexander~V.}\ \bibnamefont
  {Shapeev}}, \ and\ \bibinfo {author} {\bibfnamefont {Xiaoying}\ \bibnamefont
  {Zhuang}},\ }\bibfield  {title} {\enquote {\bibinfo {title} {Exceptional
  piezoelectricity, high thermal conductivity and stiffness and promising
  photocatalysis in two-dimensional mosi2n4 family confirmed by
  first-principles},}\ }\href {\doibase
  https://doi.org/10.1016/j.nanoen.2020.105716} {\bibfield  {journal} {\bibinfo
   {journal} {Nano Energy}\ }\textbf {\bibinfo {volume} {82}},\ \bibinfo
  {pages} {105716} (\bibinfo {year} {2021})}\BibitemShut {NoStop}%
\bibitem [{\citenamefont {Jian}\ \emph {et~al.}(2021)\citenamefont {Jian},
  \citenamefont {Ma}, \citenamefont {Zhang},\ and\ \citenamefont
  {Yong}}]{Jian}%
  \BibitemOpen
  \bibfield  {author} {\bibinfo {author} {\bibfnamefont {Chao-chao}\
  \bibnamefont {Jian}}, \bibinfo {author} {\bibfnamefont {Xiangchao}\
  \bibnamefont {Ma}}, \bibinfo {author} {\bibfnamefont {Jianqi}\ \bibnamefont
  {Zhang}}, \ and\ \bibinfo {author} {\bibfnamefont {Xin}\ \bibnamefont
  {Yong}},\ }\bibfield  {title} {\enquote {\bibinfo {title} {Strained mosi2n4
  monolayers with excellent solar energy absorption and carrier transport
  properties},}\ }\href {\doibase 10.1021/acs.jpcc.1c03585} {\bibfield
  {journal} {\bibinfo  {journal} {The Journal of Physical Chemistry C}\
  }\textbf {\bibinfo {volume} {125}},\ \bibinfo {pages} {15185--15193}
  (\bibinfo {year} {2021})}\BibitemShut {NoStop}%
\bibitem [{\citenamefont {Wang}\ \emph {et~al.}(2021)\citenamefont {Wang},
  \citenamefont {Cao}, \citenamefont {Liang}, \citenamefont {Wu}, \citenamefont
  {Wang}, \citenamefont {Lee}, \citenamefont {Ong}, \citenamefont {Yang},
  \citenamefont {Ang}, \citenamefont {Yang},\ and\ \citenamefont
  {Ang}}]{Wang2021}%
  \BibitemOpen
  \bibfield  {author} {\bibinfo {author} {\bibfnamefont {Qianqian}\
  \bibnamefont {Wang}}, \bibinfo {author} {\bibfnamefont {Liemao}\ \bibnamefont
  {Cao}}, \bibinfo {author} {\bibfnamefont {Shi-Jun}\ \bibnamefont {Liang}},
  \bibinfo {author} {\bibfnamefont {Weikang}\ \bibnamefont {Wu}}, \bibinfo
  {author} {\bibfnamefont {Guangzhao}\ \bibnamefont {Wang}}, \bibinfo {author}
  {\bibfnamefont {Ching~Hua}\ \bibnamefont {Lee}}, \bibinfo {author}
  {\bibfnamefont {Wee~Liat}\ \bibnamefont {Ong}}, \bibinfo {author}
  {\bibfnamefont {Hui~Ying}\ \bibnamefont {Yang}}, \bibinfo {author}
  {\bibfnamefont {Lay~Kee}\ \bibnamefont {Ang}}, \bibinfo {author}
  {\bibfnamefont {Shengyuan~A.}\ \bibnamefont {Yang}}, \ and\ \bibinfo {author}
  {\bibfnamefont {Yee~Sin}\ \bibnamefont {Ang}},\ }\bibfield  {title} {\enquote
  {\bibinfo {title} {Efficient ohmic contacts and built-in atomic sublayer
  protection in mosi2n4 and wsi2n4 monolayers},}\ }\href {\doibase
  10.1038/s41699-021-00251-y} {\bibfield  {journal} {\bibinfo  {journal} {npj
  2D Materials and Applications}\ }\textbf {\bibinfo {volume} {5}},\ \bibinfo
  {pages} {71} (\bibinfo {year} {2021})}\BibitemShut {NoStop}%
\bibitem [{\citenamefont {Nandan}\ \emph
  {et~al.}(2021{\natexlab{a}})\citenamefont {Nandan}, \citenamefont {Ghosh},
  \citenamefont {Agarwal}, \citenamefont {Bhowmick},\ and\ \citenamefont
  {Chauhan}}]{9646230}%
  \BibitemOpen
  \bibfield  {author} {\bibinfo {author} {\bibfnamefont {Keshari}\ \bibnamefont
  {Nandan}}, \bibinfo {author} {\bibfnamefont {Barun}\ \bibnamefont {Ghosh}},
  \bibinfo {author} {\bibfnamefont {Amit}\ \bibnamefont {Agarwal}}, \bibinfo
  {author} {\bibfnamefont {Somnath}\ \bibnamefont {Bhowmick}}, \ and\ \bibinfo
  {author} {\bibfnamefont {Yogesh~S.}\ \bibnamefont {Chauhan}},\ }\bibfield
  {title} {\enquote {\bibinfo {title} {Two-dimensional mosi2n4: An excellent
  2-d semiconductor for field-effect transistors},}\ }\href {\doibase
  10.1109/TED.2021.3130834} {\bibfield  {journal} {\bibinfo  {journal} {IEEE
  Transactions on Electron Devices}\ ,\ \bibinfo {pages} {1--8}} (\bibinfo
  {year} {2021}{\natexlab{a}})}\BibitemShut {NoStop}%
\bibitem [{\citenamefont {Wang}\ and\ \citenamefont
  {Lundstrom}(2002)}]{1175936}%
  \BibitemOpen
  \bibfield  {author} {\bibinfo {author} {\bibfnamefont {Jing}\ \bibnamefont
  {Wang}}\ and\ \bibinfo {author} {\bibfnamefont {M.}~\bibnamefont
  {Lundstrom}},\ }\bibfield  {title} {\enquote {\bibinfo {title} {Does
  source-to-drain tunneling limit the ultimate scaling of mosfets?}}\ }in\
  \href {\doibase 10.1109/IEDM.2002.1175936} {\emph {\bibinfo {booktitle}
  {Digest. International Electron Devices Meeting,}}}\ (\bibinfo {year}
  {2002})\ pp.\ \bibinfo {pages} {707--710}\BibitemShut {NoStop}%
\bibitem [{Note1()}]{Note1}%
  \BibitemOpen
  \bibinfo {note} {The Supporting information has more discussions on the i)
  structural and electronic properties of the MoA$_2$Z$_4$ series of materials,
  ii) their passivated structures, iii) and the transport properties of
  transistors with nanoribbon of MoSi$_2$N$_4$ with different
  passivations.}\BibitemShut {Stop}%
\bibitem [{\citenamefont {Wang}\ \emph {et~al.}(2020)\citenamefont {Wang},
  \citenamefont {Shi}, \citenamefont {Liu}, \citenamefont {Hong}, \citenamefont
  {Chen}, \citenamefont {Li}, \citenamefont {Gao}, \citenamefont {Ren},
  \citenamefont {Cheng}, \citenamefont {Li} \emph
  {et~al.}}]{wang2020structure}%
  \BibitemOpen
  \bibfield  {author} {\bibinfo {author} {\bibfnamefont {Lei}\ \bibnamefont
  {Wang}}, \bibinfo {author} {\bibfnamefont {Yongpeng}\ \bibnamefont {Shi}},
  \bibinfo {author} {\bibfnamefont {Mingfeng}\ \bibnamefont {Liu}}, \bibinfo
  {author} {\bibfnamefont {Yi-Lun}\ \bibnamefont {Hong}}, \bibinfo {author}
  {\bibfnamefont {Ming-Xing}\ \bibnamefont {Chen}}, \bibinfo {author}
  {\bibfnamefont {Ronghan}\ \bibnamefont {Li}}, \bibinfo {author}
  {\bibfnamefont {Qiang}\ \bibnamefont {Gao}}, \bibinfo {author} {\bibfnamefont
  {Wencai}\ \bibnamefont {Ren}}, \bibinfo {author} {\bibfnamefont {Hui-Ming}\
  \bibnamefont {Cheng}}, \bibinfo {author} {\bibfnamefont {Yiyi}\ \bibnamefont
  {Li}},  \emph {et~al.},\ }\bibfield  {title} {\enquote {\bibinfo {title}
  {Structure-driven intercalated architecture of septuple-atomic-layer $
  ma\_2z\_4 $ family with diverse properties from semiconductor to topological
  insulator to ising superconductor},}\ }\href@noop {} {\bibfield  {journal}
  {\bibinfo  {journal} {arXiv preprint arXiv:2008.02981}\ } (\bibinfo {year}
  {2020})}\BibitemShut {NoStop}%
\bibitem [{\citenamefont {Yang}\ \emph {et~al.}(2011)\citenamefont {Yang},
  \citenamefont {Shi}, \citenamefont {Zhang}, \citenamefont {Shi},\ and\
  \citenamefont {Zhang}}]{NL_H_GNR}%
  \BibitemOpen
  \bibfield  {author} {\bibinfo {author} {\bibfnamefont {Rong}\ \bibnamefont
  {Yang}}, \bibinfo {author} {\bibfnamefont {Zhiwen}\ \bibnamefont {Shi}},
  \bibinfo {author} {\bibfnamefont {Lianchang}\ \bibnamefont {Zhang}}, \bibinfo
  {author} {\bibfnamefont {Dongxia}\ \bibnamefont {Shi}}, \ and\ \bibinfo
  {author} {\bibfnamefont {Guangyu}\ \bibnamefont {Zhang}},\ }\bibfield
  {title} {\enquote {\bibinfo {title} {Observation of raman g-peak split for
  graphene nanoribbons with hydrogen-terminated zigzag edges},}\ }\href
  {\doibase 10.1021/nl201387x} {\bibfield  {journal} {\bibinfo  {journal} {Nano
  Letters}\ }\textbf {\bibinfo {volume} {11}},\ \bibinfo {pages} {4083--4088}
  (\bibinfo {year} {2011})}\BibitemShut {NoStop}%
\bibitem [{\citenamefont {Zhang}\ \emph {et~al.}(2013)\citenamefont {Zhang},
  \citenamefont {Yazyev}, \citenamefont {Feng}, \citenamefont {Xie},
  \citenamefont {Tao}, \citenamefont {Chen}, \citenamefont {Jiao},
  \citenamefont {Pedramrazi}, \citenamefont {Zettl}, \citenamefont {Louie},
  \citenamefont {Dai},\ and\ \citenamefont {Crommie}}]{NL_H_GNR_1}%
  \BibitemOpen
  \bibfield  {author} {\bibinfo {author} {\bibfnamefont {Xiaowei}\ \bibnamefont
  {Zhang}}, \bibinfo {author} {\bibfnamefont {Oleg~V.}\ \bibnamefont {Yazyev}},
  \bibinfo {author} {\bibfnamefont {Juanjuan}\ \bibnamefont {Feng}}, \bibinfo
  {author} {\bibfnamefont {Liming}\ \bibnamefont {Xie}}, \bibinfo {author}
  {\bibfnamefont {Chenggang}\ \bibnamefont {Tao}}, \bibinfo {author}
  {\bibfnamefont {Yen-Chia}\ \bibnamefont {Chen}}, \bibinfo {author}
  {\bibfnamefont {Liying}\ \bibnamefont {Jiao}}, \bibinfo {author}
  {\bibfnamefont {Zahra}\ \bibnamefont {Pedramrazi}}, \bibinfo {author}
  {\bibfnamefont {Alex}\ \bibnamefont {Zettl}}, \bibinfo {author}
  {\bibfnamefont {Steven~G.}\ \bibnamefont {Louie}}, \bibinfo {author}
  {\bibfnamefont {Hongjie}\ \bibnamefont {Dai}}, \ and\ \bibinfo {author}
  {\bibfnamefont {Michael~F.}\ \bibnamefont {Crommie}},\ }\bibfield  {title}
  {\enquote {\bibinfo {title} {Experimentally engineering the edge termination
  of graphene nanoribbons},}\ }\href {\doibase 10.1021/nn303730v} {\bibfield
  {journal} {\bibinfo  {journal} {ACS Nano}\ }\textbf {\bibinfo {volume} {7}},\
  \bibinfo {pages} {198--202} (\bibinfo {year} {2013})}\BibitemShut {NoStop}%
\bibitem [{\citenamefont {Morita}\ and\ \citenamefont
  {Tokumoto}(1995)}]{Morita}%
  \BibitemOpen
  \bibfield  {author} {\bibinfo {author} {\bibfnamefont {Yukinori}\
  \bibnamefont {Morita}}\ and\ \bibinfo {author} {\bibfnamefont {Hiroshi}\
  \bibnamefont {Tokumoto}},\ }\bibfield  {title} {\enquote {\bibinfo {title}
  {Ideal hydrogen termination of si(001) surface by wet‐chemical
  preparation},}\ }\href {\doibase 10.1063/1.114326} {\bibfield  {journal}
  {\bibinfo  {journal} {Applied Physics Letters}\ }\textbf {\bibinfo {volume}
  {67}},\ \bibinfo {pages} {2654--2656} (\bibinfo {year} {1995})}\BibitemShut
  {NoStop}%
\bibitem [{\citenamefont {Bansal}\ \emph {et~al.}(2005)\citenamefont {Bansal},
  \citenamefont {Paul},\ and\ \citenamefont {Roy}}]{1386595}%
  \BibitemOpen
  \bibfield  {author} {\bibinfo {author} {\bibfnamefont {A.}~\bibnamefont
  {Bansal}}, \bibinfo {author} {\bibfnamefont {B.C.}\ \bibnamefont {Paul}}, \
  and\ \bibinfo {author} {\bibfnamefont {K.}~\bibnamefont {Roy}},\ }\bibfield
  {title} {\enquote {\bibinfo {title} {Modeling and optimization of fringe
  capacitance of nanoscale dgmos devices},}\ }\href {\doibase
  10.1109/TED.2004.842713} {\bibfield  {journal} {\bibinfo  {journal} {IEEE
  Transactions on Electron Devices}\ }\textbf {\bibinfo {volume} {52}},\
  \bibinfo {pages} {256--262} (\bibinfo {year} {2005})}\BibitemShut {NoStop}%
\bibitem [{\citenamefont {Nandan}\ \emph
  {et~al.}(2021{\natexlab{b}})\citenamefont {Nandan}, \citenamefont {Agarwal},
  \citenamefont {Bhowmick},\ and\ \citenamefont {Chauhan}}]{Keshari_TED_PdSe2}%
  \BibitemOpen
  \bibfield  {author} {\bibinfo {author} {\bibfnamefont {Keshari}\ \bibnamefont
  {Nandan}}, \bibinfo {author} {\bibfnamefont {Amit}\ \bibnamefont {Agarwal}},
  \bibinfo {author} {\bibfnamefont {Somnath}\ \bibnamefont {Bhowmick}}, \ and\
  \bibinfo {author} {\bibfnamefont {Yogesh~S.}\ \bibnamefont {Chauhan}},\
  }\bibfield  {title} {\enquote {\bibinfo {title} {Performance investigation of
  p-{FET}s based on highly air-stable monolayer pentagonal {Pd}{Se}$_2$},}\
  }\href {\doibase 10.1109/TED.2021.3119552} {\bibfield  {journal} {\bibinfo
  {journal} {IEEE Transactions on Electron Devices}\ ,\ \bibinfo {pages}
  {1--7}} (\bibinfo {year} {2021}{\natexlab{b}})}\BibitemShut {NoStop}%
\bibitem [{\citenamefont {Lopez-Bezanilla}\ \emph {et~al.}(2012)\citenamefont
  {Lopez-Bezanilla}, \citenamefont {Huang}, \citenamefont {Terrones},\ and\
  \citenamefont {Sumpter}}]{Lopez_JPCC}%
  \BibitemOpen
  \bibfield  {author} {\bibinfo {author} {\bibfnamefont {Alejandro}\
  \bibnamefont {Lopez-Bezanilla}}, \bibinfo {author} {\bibfnamefont {Jingsong}\
  \bibnamefont {Huang}}, \bibinfo {author} {\bibfnamefont {Humberto}\
  \bibnamefont {Terrones}}, \ and\ \bibinfo {author} {\bibfnamefont {Bobby~G.}\
  \bibnamefont {Sumpter}},\ }\bibfield  {title} {\enquote {\bibinfo {title}
  {Structure and electronic properties of edge-functionalized armchair boron
  nitride nanoribbons},}\ }\href {\doibase 10.1021/jp3036583} {\bibfield
  {journal} {\bibinfo  {journal} {The Journal of Physical Chemistry C}\
  }\textbf {\bibinfo {volume} {116}},\ \bibinfo {pages} {15675--15681}
  (\bibinfo {year} {2012})}\BibitemShut {NoStop}%
\bibitem [{\citenamefont {Ren}\ \emph {et~al.}(2018)\citenamefont {Ren},
  \citenamefont {Cheng}, \citenamefont {Zhang},\ and\ \citenamefont
  {Zhou}}]{Ren2018_SR}%
  \BibitemOpen
  \bibfield  {author} {\bibinfo {author} {\bibfnamefont {Yi}~\bibnamefont
  {Ren}}, \bibinfo {author} {\bibfnamefont {Fang}\ \bibnamefont {Cheng}},
  \bibinfo {author} {\bibfnamefont {Z.~H.}\ \bibnamefont {Zhang}}, \ and\
  \bibinfo {author} {\bibfnamefont {Guanghui}\ \bibnamefont {Zhou}},\
  }\bibfield  {title} {\enquote {\bibinfo {title} {Half metal phase in the
  zigzag phosphorene nanoribbon},}\ }\href {\doibase
  10.1038/s41598-018-21294-0} {\bibfield  {journal} {\bibinfo  {journal}
  {Scientific Reports}\ }\textbf {\bibinfo {volume} {8}},\ \bibinfo {pages}
  {2932} (\bibinfo {year} {2018})}\BibitemShut {NoStop}%
\bibitem [{\citenamefont {Zhu}\ and\ \citenamefont {Su}(2010)}]{Zhu_JPCC}%
  \BibitemOpen
  \bibfield  {author} {\bibinfo {author} {\bibfnamefont {Xi}~\bibnamefont
  {Zhu}}\ and\ \bibinfo {author} {\bibfnamefont {Haibin}\ \bibnamefont {Su}},\
  }\bibfield  {title} {\enquote {\bibinfo {title} {Excitons of edge and surface
  functionalized graphene nanoribbons},}\ }\href {\doibase 10.1021/jp102341b}
  {\bibfield  {journal} {\bibinfo  {journal} {The Journal of Physical Chemistry
  C}\ }\textbf {\bibinfo {volume} {114}},\ \bibinfo {pages} {17257--17262}
  (\bibinfo {year} {2010})}\BibitemShut {NoStop}%
\bibitem [{\citenamefont {Lopez-Bezanilla}\ \emph {et~al.}(2011)\citenamefont
  {Lopez-Bezanilla}, \citenamefont {Huang}, \citenamefont {Terrones},\ and\
  \citenamefont {Sumpter}}]{Lopez_Nano}%
  \BibitemOpen
  \bibfield  {author} {\bibinfo {author} {\bibfnamefont {Alejandro}\
  \bibnamefont {Lopez-Bezanilla}}, \bibinfo {author} {\bibfnamefont {Jingsong}\
  \bibnamefont {Huang}}, \bibinfo {author} {\bibfnamefont {Humberto}\
  \bibnamefont {Terrones}}, \ and\ \bibinfo {author} {\bibfnamefont {Bobby~G.}\
  \bibnamefont {Sumpter}},\ }\bibfield  {title} {\enquote {\bibinfo {title}
  {Boron nitride nanoribbons become metallic},}\ }\href {\doibase
  10.1021/nl201616h} {\bibfield  {journal} {\bibinfo  {journal} {Nano Letters}\
  }\textbf {\bibinfo {volume} {11}},\ \bibinfo {pages} {3267--3273} (\bibinfo
  {year} {2011})}\BibitemShut {NoStop}%
\bibitem [{\citenamefont {Cervantes-Sodi}\ \emph {et~al.}(2008)\citenamefont
  {Cervantes-Sodi}, \citenamefont {Cs\'anyi}, \citenamefont {Piscanec},\ and\
  \citenamefont {Ferrari}}]{Sodi_PRB}%
  \BibitemOpen
  \bibfield  {author} {\bibinfo {author} {\bibfnamefont {F.}~\bibnamefont
  {Cervantes-Sodi}}, \bibinfo {author} {\bibfnamefont {G.}~\bibnamefont
  {Cs\'anyi}}, \bibinfo {author} {\bibfnamefont {S.}~\bibnamefont {Piscanec}},
  \ and\ \bibinfo {author} {\bibfnamefont {A.~C.}\ \bibnamefont {Ferrari}},\
  }\bibfield  {title} {\enquote {\bibinfo {title} {Edge-functionalized and
  substitutionally doped graphene nanoribbons: Electronic and spin
  properties},}\ }\href {\doibase 10.1103/PhysRevB.77.165427} {\bibfield
  {journal} {\bibinfo  {journal} {Phys. Rev. B}\ }\textbf {\bibinfo {volume}
  {77}},\ \bibinfo {pages} {165427} (\bibinfo {year} {2008})}\BibitemShut
  {NoStop}%
\bibitem [{\citenamefont {Cao}\ \emph {et~al.}(2020)\citenamefont {Cao},
  \citenamefont {Li}, \citenamefont {Li},\ and\ \citenamefont
  {Zhou}}]{D0TC01764G}%
  \BibitemOpen
  \bibfield  {author} {\bibinfo {author} {\bibfnamefont {Liemao}\ \bibnamefont
  {Cao}}, \bibinfo {author} {\bibfnamefont {Xiaobo}\ \bibnamefont {Li}},
  \bibinfo {author} {\bibfnamefont {Yun}\ \bibnamefont {Li}}, \ and\ \bibinfo
  {author} {\bibfnamefont {Guanghui}\ \bibnamefont {Zhou}},\ }\bibfield
  {title} {\enquote {\bibinfo {title} {Electrical properties and spintronic
  application of carbon phosphide nanoribbons with edge functionalization},}\
  }\href {\doibase 10.1039/D0TC01764G} {\bibfield  {journal} {\bibinfo
  {journal} {J. Mater. Chem. C}\ }\textbf {\bibinfo {volume} {8}},\ \bibinfo
  {pages} {9313--9321} (\bibinfo {year} {2020})}\BibitemShut {NoStop}%
\bibitem [{\citenamefont {Panighel}\ \emph {et~al.}(2020)\citenamefont
  {Panighel}, \citenamefont {Quiroga}, \citenamefont {Brandimarte},
  \citenamefont {Moreno}, \citenamefont {Garcia-Lekue}, \citenamefont
  {Vilas-Varela}, \citenamefont {Rey}, \citenamefont {Sauthier}, \citenamefont
  {Ceballos}, \citenamefont {Peña},\ and\ \citenamefont
  {Mugarza}}]{Mirco_Nano}%
  \BibitemOpen
  \bibfield  {author} {\bibinfo {author} {\bibfnamefont {Mirco}\ \bibnamefont
  {Panighel}}, \bibinfo {author} {\bibfnamefont {Sabela}\ \bibnamefont
  {Quiroga}}, \bibinfo {author} {\bibfnamefont {Pedro}\ \bibnamefont
  {Brandimarte}}, \bibinfo {author} {\bibfnamefont {Cesar}\ \bibnamefont
  {Moreno}}, \bibinfo {author} {\bibfnamefont {Aran}\ \bibnamefont
  {Garcia-Lekue}}, \bibinfo {author} {\bibfnamefont {Manuel}\ \bibnamefont
  {Vilas-Varela}}, \bibinfo {author} {\bibfnamefont {Dulce}\ \bibnamefont
  {Rey}}, \bibinfo {author} {\bibfnamefont {Guillaume}\ \bibnamefont
  {Sauthier}}, \bibinfo {author} {\bibfnamefont {Gustavo}\ \bibnamefont
  {Ceballos}}, \bibinfo {author} {\bibfnamefont {Diego}\ \bibnamefont {Peña}},
  \ and\ \bibinfo {author} {\bibfnamefont {Aitor}\ \bibnamefont {Mugarza}},\
  }\bibfield  {title} {\enquote {\bibinfo {title} {Stabilizing edge
  fluorination in graphene nanoribbons},}\ }\href {\doibase
  10.1021/acsnano.0c01837} {\bibfield  {journal} {\bibinfo  {journal} {ACS
  Nano}\ }\textbf {\bibinfo {volume} {14}},\ \bibinfo {pages} {11120--11129}
  (\bibinfo {year} {2020})}\BibitemShut {NoStop}%
\bibitem [{\citenamefont {Henkelman}\ \emph {et~al.}(2006)\citenamefont
  {Henkelman}, \citenamefont {Arnaldsson},\ and\ \citenamefont
  {Jónsson}}]{HENKELMAN2006354}%
  \BibitemOpen
  \bibfield  {author} {\bibinfo {author} {\bibfnamefont {Graeme}\ \bibnamefont
  {Henkelman}}, \bibinfo {author} {\bibfnamefont {Andri}\ \bibnamefont
  {Arnaldsson}}, \ and\ \bibinfo {author} {\bibfnamefont {Hannes}\ \bibnamefont
  {Jónsson}},\ }\bibfield  {title} {\enquote {\bibinfo {title} {A fast and
  robust algorithm for bader decomposition of charge density},}\ }\href
  {\doibase https://doi.org/10.1016/j.commatsci.2005.04.010} {\bibfield
  {journal} {\bibinfo  {journal} {Computational Materials Science}\ }\textbf
  {\bibinfo {volume} {36}},\ \bibinfo {pages} {354--360} (\bibinfo {year}
  {2006})}\BibitemShut {NoStop}%
\bibitem [{\citenamefont {Yu}\ and\ \citenamefont {Trinkle}(2011)}]{Bader1}%
  \BibitemOpen
  \bibfield  {author} {\bibinfo {author} {\bibfnamefont {Min}\ \bibnamefont
  {Yu}}\ and\ \bibinfo {author} {\bibfnamefont {Dallas~R.}\ \bibnamefont
  {Trinkle}},\ }\bibfield  {title} {\enquote {\bibinfo {title} {Accurate and
  efficient algorithm for bader charge integration},}\ }\href {\doibase
  10.1063/1.3553716} {\bibfield  {journal} {\bibinfo  {journal} {The Journal of
  Chemical Physics}\ }\textbf {\bibinfo {volume} {134}},\ \bibinfo {pages}
  {064111} (\bibinfo {year} {2011})}\BibitemShut {NoStop}%
\bibitem [{\citenamefont {Liu}(2020)}]{PhysRevApplied.13.064037}%
  \BibitemOpen
  \bibfield  {author} {\bibinfo {author} {\bibfnamefont {Fei}\ \bibnamefont
  {Liu}},\ }\bibfield  {title} {\enquote {\bibinfo {title} {Switching at less
  than 60 mv/decade with a ``cold'' metal as the injection source},}\ }\href
  {\doibase 10.1103/PhysRevApplied.13.064037} {\bibfield  {journal} {\bibinfo
  {journal} {Phys. Rev. Applied}\ }\textbf {\bibinfo {volume} {13}},\ \bibinfo
  {pages} {064037} (\bibinfo {year} {2020})}\BibitemShut {NoStop}%
\bibitem [{\citenamefont {Logoteta}\ \emph {et~al.}(2020)\citenamefont
  {Logoteta}, \citenamefont {Cao}, \citenamefont {Pala}, \citenamefont
  {Dollfus}, \citenamefont {Lee},\ and\ \citenamefont {Iannaccone}}]{PRR_D}%
  \BibitemOpen
  \bibfield  {author} {\bibinfo {author} {\bibfnamefont {D.}~\bibnamefont
  {Logoteta}}, \bibinfo {author} {\bibfnamefont {J.}~\bibnamefont {Cao}},
  \bibinfo {author} {\bibfnamefont {M.}~\bibnamefont {Pala}}, \bibinfo {author}
  {\bibfnamefont {P.}~\bibnamefont {Dollfus}}, \bibinfo {author} {\bibfnamefont
  {Y.}~\bibnamefont {Lee}}, \ and\ \bibinfo {author} {\bibfnamefont
  {G.}~\bibnamefont {Iannaccone}},\ }\bibfield  {title} {\enquote {\bibinfo
  {title} {Cold-source paradigm for steep-slope transistors based on van der
  waals heterojunctions},}\ }\href {\doibase 10.1103/PhysRevResearch.2.043286}
  {\bibfield  {journal} {\bibinfo  {journal} {Phys. Rev. Research}\ }\textbf
  {\bibinfo {volume} {2}},\ \bibinfo {pages} {043286} (\bibinfo {year}
  {2020})}\BibitemShut {NoStop}%
\bibitem [{\citenamefont {Logoteta}\ \emph {et~al.}(2021)\citenamefont
  {Logoteta}, \citenamefont {Cao}, \citenamefont {Pala}, \citenamefont
  {Marconcini},\ and\ \citenamefont {Iannaccone}}]{logoteta2021intrinsic}%
  \BibitemOpen
  \bibfield  {author} {\bibinfo {author} {\bibfnamefont {Demetrio}\
  \bibnamefont {Logoteta}}, \bibinfo {author} {\bibfnamefont {Jiang}\
  \bibnamefont {Cao}}, \bibinfo {author} {\bibfnamefont {Marco}\ \bibnamefont
  {Pala}}, \bibinfo {author} {\bibfnamefont {Paolo}\ \bibnamefont
  {Marconcini}}, \ and\ \bibinfo {author} {\bibfnamefont {Giuseppe}\
  \bibnamefont {Iannaccone}},\ }\bibfield  {title} {\enquote {\bibinfo {title}
  {Intrinsic subthermionic capabilities and high performance of
  easy-to-fabricate monolayer metal dihalide mosfets},}\ }\href@noop {}
  {\bibfield  {journal} {\bibinfo  {journal} {arXiv preprint arXiv:2106.12077}\
  } (\bibinfo {year} {2021})}\BibitemShut {NoStop}%
\bibitem [{\citenamefont {Giannozzi}\ \emph {et~al.}(2009)\citenamefont
  {Giannozzi}, \citenamefont {Baroni}, \citenamefont {Bonini}, \citenamefont
  {Calandra}, \citenamefont {Car}, \citenamefont {Cavazzoni}, \citenamefont
  {Ceresoli}, \citenamefont {Chiarotti}, \citenamefont {Cococcioni},
  \citenamefont {Dabo}, \citenamefont {Corso}, \citenamefont {de~Gironcoli},
  \citenamefont {Fabris}, \citenamefont {Fratesi}, \citenamefont {Gebauer},
  \citenamefont {Gerstmann}, \citenamefont {Gougoussis}, \citenamefont
  {Kokalj}, \citenamefont {Lazzeri}, \citenamefont {Martin-Samos},
  \citenamefont {Marzari}, \citenamefont {Mauri}, \citenamefont {Mazzarello},
  \citenamefont {Paolini}, \citenamefont {Pasquarello}, \citenamefont
  {Paulatto}, \citenamefont {Sbraccia}, \citenamefont {Scandolo}, \citenamefont
  {Sclauzero}, \citenamefont {Seitsonen}, \citenamefont {Smogunov},
  \citenamefont {Umari},\ and\ \citenamefont {Wentzcovitch}}]{QE_1}%
  \BibitemOpen
  \bibfield  {author} {\bibinfo {author} {\bibfnamefont {Paolo}\ \bibnamefont
  {Giannozzi}}, \bibinfo {author} {\bibfnamefont {Stefano}\ \bibnamefont
  {Baroni}}, \bibinfo {author} {\bibfnamefont {Nicola}\ \bibnamefont {Bonini}},
  \bibinfo {author} {\bibfnamefont {Matteo}\ \bibnamefont {Calandra}}, \bibinfo
  {author} {\bibfnamefont {Roberto}\ \bibnamefont {Car}}, \bibinfo {author}
  {\bibfnamefont {Carlo}\ \bibnamefont {Cavazzoni}}, \bibinfo {author}
  {\bibfnamefont {Davide}\ \bibnamefont {Ceresoli}}, \bibinfo {author}
  {\bibfnamefont {Guido~L}\ \bibnamefont {Chiarotti}}, \bibinfo {author}
  {\bibfnamefont {Matteo}\ \bibnamefont {Cococcioni}}, \bibinfo {author}
  {\bibfnamefont {Ismaila}\ \bibnamefont {Dabo}}, \bibinfo {author}
  {\bibfnamefont {Andrea~Dal}\ \bibnamefont {Corso}}, \bibinfo {author}
  {\bibfnamefont {Stefano}\ \bibnamefont {de~Gironcoli}}, \bibinfo {author}
  {\bibfnamefont {Stefano}\ \bibnamefont {Fabris}}, \bibinfo {author}
  {\bibfnamefont {Guido}\ \bibnamefont {Fratesi}}, \bibinfo {author}
  {\bibfnamefont {Ralph}\ \bibnamefont {Gebauer}}, \bibinfo {author}
  {\bibfnamefont {Uwe}\ \bibnamefont {Gerstmann}}, \bibinfo {author}
  {\bibfnamefont {Christos}\ \bibnamefont {Gougoussis}}, \bibinfo {author}
  {\bibfnamefont {Anton}\ \bibnamefont {Kokalj}}, \bibinfo {author}
  {\bibfnamefont {Michele}\ \bibnamefont {Lazzeri}}, \bibinfo {author}
  {\bibfnamefont {Layla}\ \bibnamefont {Martin-Samos}}, \bibinfo {author}
  {\bibfnamefont {Nicola}\ \bibnamefont {Marzari}}, \bibinfo {author}
  {\bibfnamefont {Francesco}\ \bibnamefont {Mauri}}, \bibinfo {author}
  {\bibfnamefont {Riccardo}\ \bibnamefont {Mazzarello}}, \bibinfo {author}
  {\bibfnamefont {Stefano}\ \bibnamefont {Paolini}}, \bibinfo {author}
  {\bibfnamefont {Alfredo}\ \bibnamefont {Pasquarello}}, \bibinfo {author}
  {\bibfnamefont {Lorenzo}\ \bibnamefont {Paulatto}}, \bibinfo {author}
  {\bibfnamefont {Carlo}\ \bibnamefont {Sbraccia}}, \bibinfo {author}
  {\bibfnamefont {Sandro}\ \bibnamefont {Scandolo}}, \bibinfo {author}
  {\bibfnamefont {Gabriele}\ \bibnamefont {Sclauzero}}, \bibinfo {author}
  {\bibfnamefont {Ari~P}\ \bibnamefont {Seitsonen}}, \bibinfo {author}
  {\bibfnamefont {Alexander}\ \bibnamefont {Smogunov}}, \bibinfo {author}
  {\bibfnamefont {Paolo}\ \bibnamefont {Umari}}, \ and\ \bibinfo {author}
  {\bibfnamefont {Renata~M}\ \bibnamefont {Wentzcovitch}},\ }\bibfield  {title}
  {\enquote {\bibinfo {title} {{QUANTUM} {ESPRESSO}: a modular and open-source
  software project for quantum simulations of materials},}\ }\href {\doibase
  10.1088/0953-8984/21/39/395502} {\bibfield  {journal} {\bibinfo  {journal}
  {Journal of Physics: Condensed Matter}\ }\textbf {\bibinfo {volume} {21}},\
  \bibinfo {pages} {395502} (\bibinfo {year} {2009})}\BibitemShut {NoStop}%
\bibitem [{\citenamefont {Giannozzi}\ \emph {et~al.}(2017)\citenamefont
  {Giannozzi}, \citenamefont {Andreussi}, \citenamefont {Brumme}, \citenamefont
  {Bunau}, \citenamefont {Nardelli}, \citenamefont {Calandra}, \citenamefont
  {Car}, \citenamefont {Cavazzoni}, \citenamefont {Ceresoli}, \citenamefont
  {Cococcioni}, \citenamefont {Colonna}, \citenamefont {Carnimeo},
  \citenamefont {Corso}, \citenamefont {de~Gironcoli}, \citenamefont {Delugas},
  \citenamefont {DiStasio}, \citenamefont {Ferretti}, \citenamefont {Floris},
  \citenamefont {Fratesi}, \citenamefont {Fugallo}, \citenamefont {Gebauer},
  \citenamefont {Gerstmann}, \citenamefont {Giustino}, \citenamefont {Gorni},
  \citenamefont {Jia}, \citenamefont {Kawamura}, \citenamefont {Ko},
  \citenamefont {Kokalj}, \citenamefont {Kü{\c{c}}ükbenli}, \citenamefont
  {Lazzeri}, \citenamefont {Marsili}, \citenamefont {Marzari}, \citenamefont
  {Mauri}, \citenamefont {Nguyen}, \citenamefont {Nguyen}, \citenamefont {de-la
  Roza}, \citenamefont {Paulatto}, \citenamefont {Ponc{\'{e}}}, \citenamefont
  {Rocca}, \citenamefont {Sabatini}, \citenamefont {Santra}, \citenamefont
  {Schlipf}, \citenamefont {Seitsonen}, \citenamefont {Smogunov}, \citenamefont
  {Timrov}, \citenamefont {Thonhauser}, \citenamefont {Umari}, \citenamefont
  {Vast}, \citenamefont {Wu},\ and\ \citenamefont {Baroni}}]{QE_2}%
  \BibitemOpen
  \bibfield  {author} {\bibinfo {author} {\bibfnamefont {P}~\bibnamefont
  {Giannozzi}}, \bibinfo {author} {\bibfnamefont {O}~\bibnamefont {Andreussi}},
  \bibinfo {author} {\bibfnamefont {T}~\bibnamefont {Brumme}}, \bibinfo
  {author} {\bibfnamefont {O}~\bibnamefont {Bunau}}, \bibinfo {author}
  {\bibfnamefont {M~Buongiorno}\ \bibnamefont {Nardelli}}, \bibinfo {author}
  {\bibfnamefont {M}~\bibnamefont {Calandra}}, \bibinfo {author} {\bibfnamefont
  {R}~\bibnamefont {Car}}, \bibinfo {author} {\bibfnamefont {C}~\bibnamefont
  {Cavazzoni}}, \bibinfo {author} {\bibfnamefont {D}~\bibnamefont {Ceresoli}},
  \bibinfo {author} {\bibfnamefont {M}~\bibnamefont {Cococcioni}}, \bibinfo
  {author} {\bibfnamefont {N}~\bibnamefont {Colonna}}, \bibinfo {author}
  {\bibfnamefont {I}~\bibnamefont {Carnimeo}}, \bibinfo {author} {\bibfnamefont
  {A~Dal}\ \bibnamefont {Corso}}, \bibinfo {author} {\bibfnamefont
  {S}~\bibnamefont {de~Gironcoli}}, \bibinfo {author} {\bibfnamefont
  {P}~\bibnamefont {Delugas}}, \bibinfo {author} {\bibfnamefont {R~A}\
  \bibnamefont {DiStasio}}, \bibinfo {author} {\bibfnamefont {A}~\bibnamefont
  {Ferretti}}, \bibinfo {author} {\bibfnamefont {A}~\bibnamefont {Floris}},
  \bibinfo {author} {\bibfnamefont {G}~\bibnamefont {Fratesi}}, \bibinfo
  {author} {\bibfnamefont {G}~\bibnamefont {Fugallo}}, \bibinfo {author}
  {\bibfnamefont {R}~\bibnamefont {Gebauer}}, \bibinfo {author} {\bibfnamefont
  {U}~\bibnamefont {Gerstmann}}, \bibinfo {author} {\bibfnamefont
  {F}~\bibnamefont {Giustino}}, \bibinfo {author} {\bibfnamefont
  {T}~\bibnamefont {Gorni}}, \bibinfo {author} {\bibfnamefont {J}~\bibnamefont
  {Jia}}, \bibinfo {author} {\bibfnamefont {M}~\bibnamefont {Kawamura}},
  \bibinfo {author} {\bibfnamefont {H-Y}\ \bibnamefont {Ko}}, \bibinfo {author}
  {\bibfnamefont {A}~\bibnamefont {Kokalj}}, \bibinfo {author} {\bibfnamefont
  {E}~\bibnamefont {Kü{\c{c}}ükbenli}}, \bibinfo {author} {\bibfnamefont
  {M}~\bibnamefont {Lazzeri}}, \bibinfo {author} {\bibfnamefont
  {M}~\bibnamefont {Marsili}}, \bibinfo {author} {\bibfnamefont
  {N}~\bibnamefont {Marzari}}, \bibinfo {author} {\bibfnamefont
  {F}~\bibnamefont {Mauri}}, \bibinfo {author} {\bibfnamefont {N~L}\
  \bibnamefont {Nguyen}}, \bibinfo {author} {\bibfnamefont {H-V}\ \bibnamefont
  {Nguyen}}, \bibinfo {author} {\bibfnamefont {A~Otero}\ \bibnamefont {de-la
  Roza}}, \bibinfo {author} {\bibfnamefont {L}~\bibnamefont {Paulatto}},
  \bibinfo {author} {\bibfnamefont {S}~\bibnamefont {Ponc{\'{e}}}}, \bibinfo
  {author} {\bibfnamefont {D}~\bibnamefont {Rocca}}, \bibinfo {author}
  {\bibfnamefont {R}~\bibnamefont {Sabatini}}, \bibinfo {author} {\bibfnamefont
  {B}~\bibnamefont {Santra}}, \bibinfo {author} {\bibfnamefont {M}~\bibnamefont
  {Schlipf}}, \bibinfo {author} {\bibfnamefont {A~P}\ \bibnamefont
  {Seitsonen}}, \bibinfo {author} {\bibfnamefont {A}~\bibnamefont {Smogunov}},
  \bibinfo {author} {\bibfnamefont {I}~\bibnamefont {Timrov}}, \bibinfo
  {author} {\bibfnamefont {T}~\bibnamefont {Thonhauser}}, \bibinfo {author}
  {\bibfnamefont {P}~\bibnamefont {Umari}}, \bibinfo {author} {\bibfnamefont
  {N}~\bibnamefont {Vast}}, \bibinfo {author} {\bibfnamefont {X}~\bibnamefont
  {Wu}}, \ and\ \bibinfo {author} {\bibfnamefont {S}~\bibnamefont {Baroni}},\
  }\bibfield  {title} {\enquote {\bibinfo {title} {Advanced capabilities for
  materials modelling with quantum {ESPRESSO}},}\ }\href {\doibase
  10.1088/1361-648x/aa8f79} {\bibfield  {journal} {\bibinfo  {journal} {Journal
  of Physics: Condensed Matter}\ }\textbf {\bibinfo {volume} {29}},\ \bibinfo
  {pages} {465901} (\bibinfo {year} {2017})}\BibitemShut {NoStop}%
\bibitem [{\citenamefont {Perdew}\ \emph {et~al.}(1996)\citenamefont {Perdew},
  \citenamefont {Burke},\ and\ \citenamefont {Ernzerhof}}]{PBE}%
  \BibitemOpen
  \bibfield  {author} {\bibinfo {author} {\bibfnamefont {John~P.}\ \bibnamefont
  {Perdew}}, \bibinfo {author} {\bibfnamefont {Kieron}\ \bibnamefont {Burke}},
  \ and\ \bibinfo {author} {\bibfnamefont {Matthias}\ \bibnamefont
  {Ernzerhof}},\ }\bibfield  {title} {\enquote {\bibinfo {title} {Generalized
  gradient approximation made simple},}\ }\href {\doibase
  10.1103/PhysRevLett.77.3865} {\bibfield  {journal} {\bibinfo  {journal}
  {Phys. Rev. Lett.}\ }\textbf {\bibinfo {volume} {77}},\ \bibinfo {pages}
  {3865--3868} (\bibinfo {year} {1996})}\BibitemShut {NoStop}%
\bibitem [{\citenamefont {Mostofi}\ \emph {et~al.}(2014)\citenamefont
  {Mostofi}, \citenamefont {Yates}, \citenamefont {Pizzi}, \citenamefont {Lee},
  \citenamefont {Souza}, \citenamefont {Vanderbilt},\ and\ \citenamefont
  {Marzari}}]{W90}%
  \BibitemOpen
  \bibfield  {author} {\bibinfo {author} {\bibfnamefont {Arash~A.}\
  \bibnamefont {Mostofi}}, \bibinfo {author} {\bibfnamefont {Jonathan~R.}\
  \bibnamefont {Yates}}, \bibinfo {author} {\bibfnamefont {Giovanni}\
  \bibnamefont {Pizzi}}, \bibinfo {author} {\bibfnamefont {Young-Su}\
  \bibnamefont {Lee}}, \bibinfo {author} {\bibfnamefont {Ivo}\ \bibnamefont
  {Souza}}, \bibinfo {author} {\bibfnamefont {David}\ \bibnamefont
  {Vanderbilt}}, \ and\ \bibinfo {author} {\bibfnamefont {Nicola}\ \bibnamefont
  {Marzari}},\ }\bibfield  {title} {\enquote {\bibinfo {title} {An updated
  version of wannier90: A tool for obtaining maximally-localised wannier
  functions},}\ }\href {\doibase https://doi.org/10.1016/j.cpc.2014.05.003}
  {\bibfield  {journal} {\bibinfo  {journal} {Computer Physics Communications}\
  }\textbf {\bibinfo {volume} {185}},\ \bibinfo {pages} {2309 -- 2310}
  (\bibinfo {year} {2014})}\BibitemShut {NoStop}%
\bibitem [{\citenamefont {Datta}(2005)}]{datta2005quantum}%
  \BibitemOpen
  \bibfield  {author} {\bibinfo {author} {\bibfnamefont {Supriyo}\ \bibnamefont
  {Datta}},\ }\href@noop {} {\emph {\bibinfo {title} {Quantum transport: atom
  to transistor}}}\ (\bibinfo  {publisher} {Cambridge university press},\
  \bibinfo {year} {2005})\BibitemShut {NoStop}%
\bibitem [{\citenamefont {Bruzzone}\ \emph {et~al.}(2014)\citenamefont
  {Bruzzone}, \citenamefont {Iannaccone}, \citenamefont {Marzari},\ and\
  \citenamefont {Fiori}}]{NanoTCAD}%
  \BibitemOpen
  \bibfield  {author} {\bibinfo {author} {\bibfnamefont {S.}~\bibnamefont
  {Bruzzone}}, \bibinfo {author} {\bibfnamefont {G.}~\bibnamefont
  {Iannaccone}}, \bibinfo {author} {\bibfnamefont {N.}~\bibnamefont {Marzari}},
  \ and\ \bibinfo {author} {\bibfnamefont {G.}~\bibnamefont {Fiori}},\
  }\bibfield  {title} {\enquote {\bibinfo {title} {An open-source multiscale
  framework for the simulation of nanoscale devices},}\ }\href {\doibase
  10.1109/TED.2013.2291909} {\bibfield  {journal} {\bibinfo  {journal} {IEEE
  Transactions on Electron Devices}\ }\textbf {\bibinfo {volume} {61}},\
  \bibinfo {pages} {48--53} (\bibinfo {year} {2014})}\BibitemShut {NoStop}%
\end{thebibliography}%
	
\end{document}